\documentclass[aps,prc,superscriptaddress,showpacs,nofootinbib,floatfix,onecolumn]{revtex4}
\usepackage{afterpage}
\usepackage{setspace}
\usepackage{rotating}
\usepackage{amssymb}
\usepackage{float}
\usepackage{harpoon}
\usepackage{amsmath}
\usepackage{atlasphysics} 
\usepackage[colorlinks,linkcolor=blue,citecolor=blue,filecolor=black,urlcolor=blue,pdfauthor={The ATLAS Collaboration}]{hyperref}
\usepackage{preprintcover}  
\PreprintCoverPaperTitle{\boldmath Measurement of long-range pseudorapidity correlations and azimuthal harmonics in $\sqrt{s_{\mathrm{NN}}}$=5.02 TeV proton-lead collisions with the ATLAS detector}
\PreprintIdNumber{CERN-PH-EP-2014-201}  

\PreprintCoverAbstract{Measurements of two-particle correlation functions and the first five azimuthal harmonics, $v_1$ to $v_5$, are presented, using 28~$\mathrm{nb}^{-1}$ of $p$+Pb collisions at a nucleon-nucleon center-of-mass energy of $\sqrt{s_{\mathrm{NN}}}=5.02$~TeV measured with the ATLAS detector at the LHC. Significant long-range ``ridge-like'' correlations are observed for pairs with small relative azimuthal angle ($|\Delta\phi|<\pi/3$) and back-to-back pairs ($|\Delta\phi|>2\pi/3$) over the transverse momentum range $0.4<p_{\rm T}<12$ GeV and in different intervals of event activity. The event activity is defined by either the number of reconstructed tracks or the total transverse energy on the Pb-fragmentation side. The azimuthal structure of such long-range correlations is Fourier decomposed to obtain the harmonics $v_n$ as a function of $p_{\rm T}$ and event activity. The extracted $v_n$ values for $n=2$ to 5 decrease with $n$. The $v_2$ and $v_3$ values are found to be positive in the measured $p_{\rm T}$ range. The $v_1$ is also measured as a function of $p_{\rm T}$ and is observed to change sign around $p_{\rm T}\approx 1.5$--2.0 GeV and then increase to about 0.1 for $p_{\rm T}>4$ GeV. The $v_2(p_{\rm T})$, $v_3(p_{\rm T})$ and $v_4(p_{\rm T})$ are compared to the $v_n$ coefficients in Pb+Pb collisions at $\sqrt{s_{\mathrm{NN}}} =2.76$~TeV with similar event multiplicities. Reasonable agreement is observed after accounting for the difference in the average $p_{\rm T}$ of particles produced in the two collision systems.}
\PreprintJournalName{Phys. Rev. C.}  

\begin{document}
\newcommand{\pp}{\mbox{$p$+$p$}}
\newcommand{\pPb}{\mbox{$p$+Pb}}
\newcommand{\ptref}{\mbox{$p_{\mathrm{T}}^{\mathrm{b}}$}}
\newcommand{\pta}{\mbox{$p_{\mathrm{T}}^{\mathrm{a}}$}}
\newcommand{\ptb}{\mbox{$p_{\mathrm{T}}^{\mathrm{b}}$}}
\newcommand{\ptab}{\mbox{$p_{\mathrm{T}}^{\mathrm{a,b}}$}}
\newcommand{\pA}{\mbox{$p$+A}}
\newcommand{\PbPb}{\mbox{Pb+Pb}}
\newcommand{\invnb}{\mbox{${\rm nb}^{-1}$}}
\newcommand{\nchave}{\mbox{$\langle N_{\mathrm{ch}}\rangle$}}
\newcommand{\signch}{\mbox{$\sigma_{\tiny{N_{\mathrm{ch}}}}$}}
\newcommand{\nchrecave}{\mbox{$\langle N_{\mathrm{ch}}^{\mathrm{rec}}\rangle$}}
\newcommand{\vndir}{\mbox{$v_{n}^{\mathrm{unsub}}$}}
\newcommand{\nch}{\mbox{$N_{\mathrm{ch}}$}}
\newcommand{\nchrec}{\mbox{$N_{\mathrm{ch}}^{\mathrm{rec}}$}}
\newcommand{\nchrecs}{\mbox{\footnotesize$N_{\mathrm{ch}}^{\mathrm{rec}}$\normalsize}}
\newcommand{\Ntrk}{\mbox{$N_{\mathrm{ch}}$}}
\newcommand{\Nevt}{\mbox{$N_{\mathrm{evt}}$}}
\newcommand{\zvtx}{\mbox{$z_{\mathrm{vtx}}$}}
\newcommand{\Npair}{\mbox{$N_{\mathrm{pair}}$}}
\newcommand{\npart}{\mbox{$N_{\mathrm{part}}$}}
\newcommand{\ETfcals}{\mbox{\footnotesize$ E_{\mathrm{T}}^{\mathrm{FCal_{Pb}}}$\normalsize}}
\newcommand{\ETfcal}{\mbox{$E_{\mathrm{T}}^{{\scriptscriptstyle \mathrm{Pb}}}$}}
\newcommand{\ETfcalave}{\mbox{$\langle E_{\mathrm{T}}^{{\scriptscriptstyle \mathrm{Pb}}}\rangle$}}
\newcommand{\ettrig}{\mbox{$E_{\mathrm{T}}^{_{\mathrm{L1}}}$}}
\newcommand{\ntrktrig}{\mbox{$N_{\mathrm{trk}}^{_{\mathrm{HLT}}}$}}
\newcommand{\eff}{\mbox{$\epsilon_{\mathrm{trk}}$}}
\newcommand{\sqrtsNN}{\mbox{$\sqrt{s_{\mathrm{NN}}}$}}
\newcommand{\sqrtsnn}{\mbox{$\sqrt{s_{\mathrm{NN}}}$}}
\newcommand{\Dphi}{\mbox{$\Delta \phi$}}
\newcommand{\Deta}{\mbox{$\Delta \eta$}}
\newcommand{\pTch}{\mbox{$p_{{\mathrm{T}}}^{\mathrm{ch}}$}}
\newcommand{\Yphi}{\mbox{$Y(\Delta\phi)$}}
\newcommand{\Yint}{\mbox{$Y_{\mathrm{int}}$}}
\newcommand{\DelYint}{\mbox{$\Delta Y_{\mathrm{int}}$}}
\newcommand{\Yintc}{\mbox{$Y_{\mathrm{int}}^{\mathrm{central}}$}}
\newcommand{\Yintp}{\mbox{$Y_{\mathrm{int}}^{\mathrm{peri}}$}}
\title{Measurement of long-range pseudorapidity correlations and azimuthal harmonics in $\sqrt{s_{\mathrm{NN}}}$=5.02 TeV proton-lead collisions with the ATLAS detector}
\author{The ATLAS Collaboration}
\begin{abstract}
Measurements of two-particle correlation functions and the first five azimuthal harmonics, $v_1$ to $v_5$, are presented, using 28~$\mathrm{nb}^{-1}$ of $p$+Pb collisions at a nucleon-nucleon center-of-mass energy of $\sqrt{s_{\mathrm{NN}}}=5.02$~TeV measured with the ATLAS detector at the LHC. Significant long-range ``ridge-like'' correlations are observed for pairs with small relative azimuthal angle ($|\Delta\phi|<\pi/3$) and back-to-back pairs ($|\Delta\phi|>2\pi/3$) over the transverse momentum range $0.4<p_{\rm T}<12$ GeV and in different intervals of event activity. The event activity is defined by either the number of reconstructed tracks or the total transverse energy on the Pb-fragmentation side. The azimuthal structure of such long-range correlations is Fourier decomposed to obtain the harmonics $v_n$ as a function of $p_{\rm T}$ and event activity. The extracted $v_n$ values for $n=2$ to 5 decrease with $n$. The $v_2$ and $v_3$ values are found to be positive in the measured $p_{\rm T}$ range. The $v_1$ is also measured as a function of $p_{\rm T}$ and is observed to change sign around $p_{\rm T}\approx 1.5$--2.0 GeV and then increase to about 0.1 for $p_{\rm T}>4$ GeV. The $v_2(p_{\rm T})$, $v_3(p_{\rm T})$ and $v_4(p_{\rm T})$ are compared to the $v_n$ coefficients in Pb+Pb collisions at $\sqrt{s_{\mathrm{NN}}} =2.76$~TeV with similar event multiplicities. Reasonable agreement is observed after accounting for the difference in the average $p_{\rm T}$ of particles produced in the two collision systems.
\end{abstract}
\maketitle
\section{Introduction}
One striking observation in high-energy nucleus-nucleus (A+A) collisions is the large anisotropy of particle production in the azimuthal angle $\phi$~\cite{Ackermann:2000tr,ALICE:2011ab}. This anisotropy is often studied via a two-particle correlation of particle pairs in relative pseudorapidity ($\Delta\eta$) and azimuthal angle ($\Delta\phi$)~\cite{Jia:2004sw,Adare:2008ae}. The anisotropy manifests itself as a strong excess of pairs at $\Delta\phi\sim0$ and $\pi$, and the magnitude of the excess is relatively constant out to large $|\Delta\eta|$~\cite{Alver:2009id,Aggarwal:2010rf,Aamodt:2011by,CMS:2012wg,Aad:2012bu}. The azimuthal structure of this ``ridge-like'' correlation is commonly characterized by its Fourier harmonics, $dN_{\mathrm{pairs}}/d\Delta\phi \sim 1 + 2\sum_{n} v_{n}^2 \cos n \Delta\phi$. While the elliptic flow, $v_2$, and triangular flow, $v_3$, are the dominant harmonics in A+A collisions, significant $v_1$, $v_4$, $v_5$ and $v_6$ harmonics have also been measured~\cite{CMS:2012wg,Aad:2012bu,Aad:2013xma,Chatrchyan:2013kba,Adamczyk:2013waa,Adare:2011tg}. These $v_n$ values are commonly interpreted as the collective hydrodynamic response of the created matter to the collision geometry and its fluctuations in the initial state~\cite{Alver:2010gr}. The success of hydrodynamic models in describing the anisotropy of particle production in heavy-ion collisions at RHIC and the LHC places important constraints on the transport properties of the produced matter~\cite{Luzum:2012wu,Teaney:2010vd,Gale:2012rq,Niemi:2012aj,Qiu:2012uy,Teaney:2013dta}.

For a small collision system, such as proton-proton ($\pp$) or proton-nucleus ($\pA$) collisions, it was assumed that the transverse size of the produced system is too small for the hydrodynamic flow description to be applicable. Thus, it came as a surprise that ridge-like structures were also observed in two-particle correlations in high-multiplicity $\pp$~\cite{Khachatryan:2010gv} and proton-lead ($\pPb$)~\cite{CMS:2012qk,Abelev:2012ola,Aad:2012gla} collisions at the LHC and later in deuteron-gold collisions~\cite{Adare:2013piz} at RHIC. A Fourier decomposition technique has been employed to study the azimuthal distribution of the ridge in $\pPb$ collisions. The transverse momentum ($\pT$) dependence of the extracted $v_2$ and $v_3$~\cite{Abelev:2012ola,Aad:2012gla}, and the particle-mass dependence of $v_2$~\cite{ABELEV:2013wsa} are found to be similar to those measured in A+A collisions. Large $v_2$ coefficients are also measured via the four-particle cumulant method~\cite{Aad:2013fja,Chatrchyan:2013nka}, suggesting that the ridge reflects genuine multi-particle correlations.

The interpretation of the long-range correlations in high-multiplicity $\pp$ and $\pPb$ collisions is currently a subject of intense study. References~\cite{Bozek:2011if,Shuryak:2013ke,Bozek:2013uha,Qin:2013bha} argue that the produced matter in these collisions is sufficiently voluminous and dense such that the hydrodynamic model framework may still apply. On the other hand, models based on gluon saturation and color connections suggest that the long-range correlations are an initial-state effect, intrinsic to QCD at high gluon density~\cite{Dusling:2012cg,Dusling:2012wy,Kovchegov:2012nd,McLerran:2013oju,Dusling:2013oia}. Recently a hybrid model that takes into account both the initial- and final-state effects has been proposed~\cite{Bzdak:2013zma}. All these approaches can describe, qualitatively and even quantitatively, the $v_2$ and $v_3$ data in the $\pPb$ collisions. 

In order to provide more insights into the nature of the ridge correlation and to discriminate between different theoretical interpretations, this paper provides a detailed measurement of the two-particle correlation and $v_n$ coefficients in $\pPb$ collisions at a nucleon-nucleon center-of-mass energy of $\sqrt{s_{_{\mathrm{NN}}}}=5.02$~TeV. The data correspond to an integrated luminosity of approximately 28~$\mathrm{nb}^{-1}$, recorded in 2013 by the ATLAS experiment at the LHC. This measurement benefits from a dedicated high-multiplicity trigger (see Sec.~\ref{sec:trig}) that provides a large sample of high-multiplicity events, not only extending the previous $v_2$ and $v_3$ results to higher $\pT$, but also enabling the first measurement of $v_1$, $v_4$ and $v_5$. The results are extracted independently for two different event-activity definitions: the total transverse energy in the forward calorimeter on the Pb-fragmentation side\footnote{ATLAS uses a right-handed coordinate system with its origin at the nominal interaction point (IP) in the center of the detector and the $z$-axis along the beam pipe. The $x$-axis points from the IP towards the center of the LHC ring, and the $y$-axis completes the right-handed system. Cylindrical coordinates $(r,\phi)$ are used in the transverse plane, $\phi$ being the azimuthal angle around the beam pipe. The pseudorapidity is defined in terms of the polar angle $\theta$ as $\eta=-\ln\tan(\theta/2)$. During 2013 $\pPb$ data taking, the beam directions were reversed approximately half-way through the running period, but in presenting results the direction of the proton beam is always chosen to point to positive $\eta$.} ($-4.9<\eta<-3.2$), $\ETfcal$, or the number of reconstructed tracks in $|\eta|<2.5$, $\nchrec$. The results are also compared to the Pb+Pb data with similar multiplicity. The analysis technique follows closely the previous ATLAS study of $v_2$ and $v_3$ based on a much smaller dataset from a short $\pPb$ run in 2012~\cite{Aad:2012gla}. 

\section{Experimental setup}
\subsection{Detector and dataset}
The ATLAS detector~\cite{Aad:2008zzm} provides nearly full solid-angle coverage of the collision point with tracking detectors, calorimeters and muon chambers. The measurements presented in this paper are performed using the ATLAS inner detector (ID), forward calorimeters (FCal), minimum-bias trigger scintillators (MBTS), zero-degree calorimeter (ZDC) and the trigger and data acquisition systems. The ID measures charged particles within the pseudorapidity region $|\eta| < 2.5$ using a combination of silicon pixel detector, silicon microstrip detector (SCT), and a straw-tube transition radiation tracker, all immersed in a 2 T axial magnetic field. The MBTS detects charged particles over 2.1 $< |\eta| <$ 3.9 using two hodoscopes of 16 counters positioned at $z$ = $\pm$3.6 m. The FCal consists of two sections that cover 3.2 $< |\eta| <$ 4.9. The FCal modules are composed of tungsten and copper absorbers with liquid argon as the active medium, which provide 10 interaction lengths of material. The ZDC, situated at $\approx$ 140~m from the collision vertex, detects neutral particles, mostly neutrons and photons, with $|\eta| >$ 8.3.

This analysis uses approximately 28~$\invnb$ of $\pPb$ data recorded by the ATLAS experiment at the LHC in 2013. The LHC was configured with a 4~\TeV\ proton beam and a 1.57~\TeV\ per-nucleon Pb beam that together produced collisions at $\sqrtsnn = 5.02$~TeV. The beam directions were reversed approximately half-way through the running period. The higher energy of the proton beam results in a net rapidity shift of the nucleon--nucleon center-of-mass frame relative to the ATLAS rest frame. This rapidity shift is $0.47$ towards the proton beam direction.

\subsection{Trigger}
\label{sec:trig}
The minimum-bias (MB) Level-1 (L1) trigger~\cite{Aad:2012xs} used for this analysis requires a signal in at least one MBTS counter on each side, or a signal in the ZDC on the Pb-fragmentation side with the trigger threshold set just below the peak corresponding to a single neutron. A timing requirement based on signals from each side of the MBTS is imposed to suppress beam backgrounds. Due to the high event rate during the run, only a small fraction of the MB events ($\sim$1/1000) were recorded. In order to enhance the number of events with high multiplicity, a dedicated high-multiplicity trigger (HMT) was implemented, which uses the ATLAS L1 and high-level trigger (HLT) systems~\cite{pPb2013}. At L1, the total transverse energy $\ettrig$ in the FCal rapidity interval is required to be above a certain threshold. In the HLT, the charged-particle tracks are reconstructed by requiring at least two hits in the pixel detector and three hits in the SCT. For each event, the collision vertex reconstructed with the highest number of online tracks is selected, and the number of tracks ($\ntrktrig$) associated with this vertex with $\pT>0.4$ GeV and a distance of closest approach of less than 4~mm is calculated. 

The HMT triggers are implemented by requiring different thresholds on the values of $\ettrig$ and $\ntrktrig$ with prescale factors adjusted to the instantaneous luminosity provided by the LHC~\cite{pPb2013}. This analysis uses the six pairs of thresholds on $\ettrig$ and $\ntrktrig$ listed in Table~\ref{tab:trig}. The $\ntrktrig \geq$ 225 trigger was not prescaled throughout the entire running period. 
\begin{table}[!h]
\centering
\begin{tabular}{|c|c|c|c|c|c|c|}\hline
$\ntrktrig$        & $\geq$100 & $\geq$130& $\geq$150& $\geq$180& $\geq$200& $\geq$225\tabularnewline\hline
$\ettrig$ [GeV]    & $\geq$10 & $\geq$10 & $\geq$50 & $\geq$50 & $\geq$65 & $\geq$65  \tabularnewline\hline
\end{tabular}
\caption{\label{tab:trig} A list of thresholds in $\ettrig$ and $\ntrktrig$ for the high-multiplicity triggers used in this analysis.}
\end{table}
\pagebreak
\section{Data analysis}
\label{sec:ana}
\subsection{Event and track selections}
\label{sec:sel}
In the offline analysis, $\pPb$ events are required to have a reconstructed vertex containing at least two associated offline tracks, with its $z$ position satisfying $|\zvtx|< 150$~mm. Non-collision backgrounds and photonuclear interactions are suppressed by requiring at least one hit in a MBTS counter on each side of the interaction point, and the difference between times measured on the two sides to be less than 10~ns. In the 2013 $\pPb$ run, the luminosity conditions provided by the LHC result in an average probability of 3\% that an event contains two or more $\pPb$ collisions (pileup). The pileup events are suppressed by rejecting events containing more than one good reconstructed vertex. The remaining pileup events are further suppressed based on the signal in the ZDC on the Pb-fragmentation side. This signal is calibrated to the number of detected neutrons ($N_{n}$) based on the location of the peak corresponding to a single neutron. The distribution of $N_{n}$ in events with pileup is broader than that for the events without pileup. Hence a simple cut on the high tail-end of the ZDC signal distribution further suppresses the pileup, while retaining more than 98\% of the events without pileup. After this pileup rejection procedure, the residual pileup fraction is estimated to be $\lesssim 10^{-2}$ in the event class with the highest track multiplicity studied in this analysis. About 57 million MB-selected events and 15 million HMT-selected events are included in this analysis.

Charged-particle tracks are reconstructed in the ID using an algorithm optimized for $\pp$ minimum-bias measurements~\cite{Aad:2010ac}: the tracks are required to have $\pT >$ 0.3 GeV and $|\eta| <$ 2.5, at least seven hits in the pixel detector and the SCT, and a hit in the first pixel layer when one is expected. In addition, the transverse ($d_0$) and longitudinal ($z_0$ $\sin\theta$) impact parameters of the track relative to the vertex are required to be less than 1.5 mm. They are also required to satisfy $|d_0/\sigma_{d_0}| <$ 3 and $|z_0\sin\theta/\sigma_z | <$ 3, respectively, where $\sigma_{d_0}$ and $\sigma_z$ are uncertainties on $d_0$ and $z_0\sin\theta$ obtained from the track-fit covariance matrix.

The efficiency, $\epsilon(\pT, \eta)$, for track reconstruction and track selection cuts is obtained using $\pPb$ Monte Carlo events produced with version 1.38b of the HIJING event generator~\cite{Wang:1991} with a center-of-mass boost matching the beam conditions. The response of the detector is simulated using GEANT4~\cite{Agostinelli:2002hh,Aad:2010ah} and the resulting events are reconstructed with the same algorithms as applied to the data. The efficiency increases with $\pT$ by 6\% between 0.3 and 0.5 GeV, and varies only weakly for $\pT >$ 0.5 GeV, where it ranges from 82\% at $\eta$ = 0 to 70\% at $|\eta|$ = 2 and 60\% at $|\eta| >$ 2.4. The efficiency is also found to vary by less than 2\% over the multiplicity range used in the analysis. The extracted efficiency function $\epsilon(\pT, \eta)$ is used in the correlation analysis, as well as to estimate the average efficiency-corrected charged-particle multiplicity in the collisions. 
\subsection{Characterization of the event activity}
The two-particle correlation (2PC) analyses are performed in event classes with different overall activity. The event activity is characterized by either $\ETfcal$, the sum of transverse energy measured on the Pb-fragmentation side of the FCal with $-4.9<\eta<-3.2$, or $\nchrec$, the offline-reconstructed track multiplicity in the ID with $|\eta|<2.5$ and $\pT>0.4$ GeV. These event-activity definitions have been used in previous $\pPb$ analyses~\cite{Aad:2012gla,Aad:2013fja,Khachatryan:2010gv,CMS:2012qk,Chatrchyan:2013nka}. Events with larger activity have on average a larger number of participating nucleons in the Pb nucleus and a smaller impact parameter. Hence the term ``centrality'', familiar in A+A collisions, is used to refer to the event activity. The terms ``central'' and ``peripheral'' are used to refer to events with large activity and small activity, respectively.

Due to the wide range of trigger thresholds and the prescale values required by the HMT triggers, the $\ETfcal$ and $\nchrec$ distributions are very different for the HMT events and the MB events. In order to properly include the HMT events in the event-activity classification, an event-by-event weight, $w=1/P$, is utilized. The combined probability, $P$, for a given event to be accepted by the MB trigger or any of the HMT triggers is calculated via the inclusion-exclusion principle as
\begin{eqnarray} 
\label{eq:setup}
P = \sum_{1\le i\le N}p_i - \sum_{1\le i<j\le N}p_ip_j+ \sum_{1\le i<j<k\le N}p_ip_jp_k-...\;\;,
\end{eqnarray}
where $N$ is the total number of triggers, and $p_i$ is the probability for the $i^{\mathrm{th}}$-trigger to accept the event, defined as zero if the event does not fire the trigger and otherwise as the inverse of the prescale factor of the trigger. The higher-order terms in Eq.~(\ref{eq:setup}) account for the probability of more than one trigger being fired. The weight factor, $w$, is calculated and applied event by event. The distribution for all events after re-weighting has the same shape as the distribution for MB events, as should be the case if the re-weighting is done correctly.

\begin{figure}[!t]
\centering
\includegraphics[width=1\columnwidth]{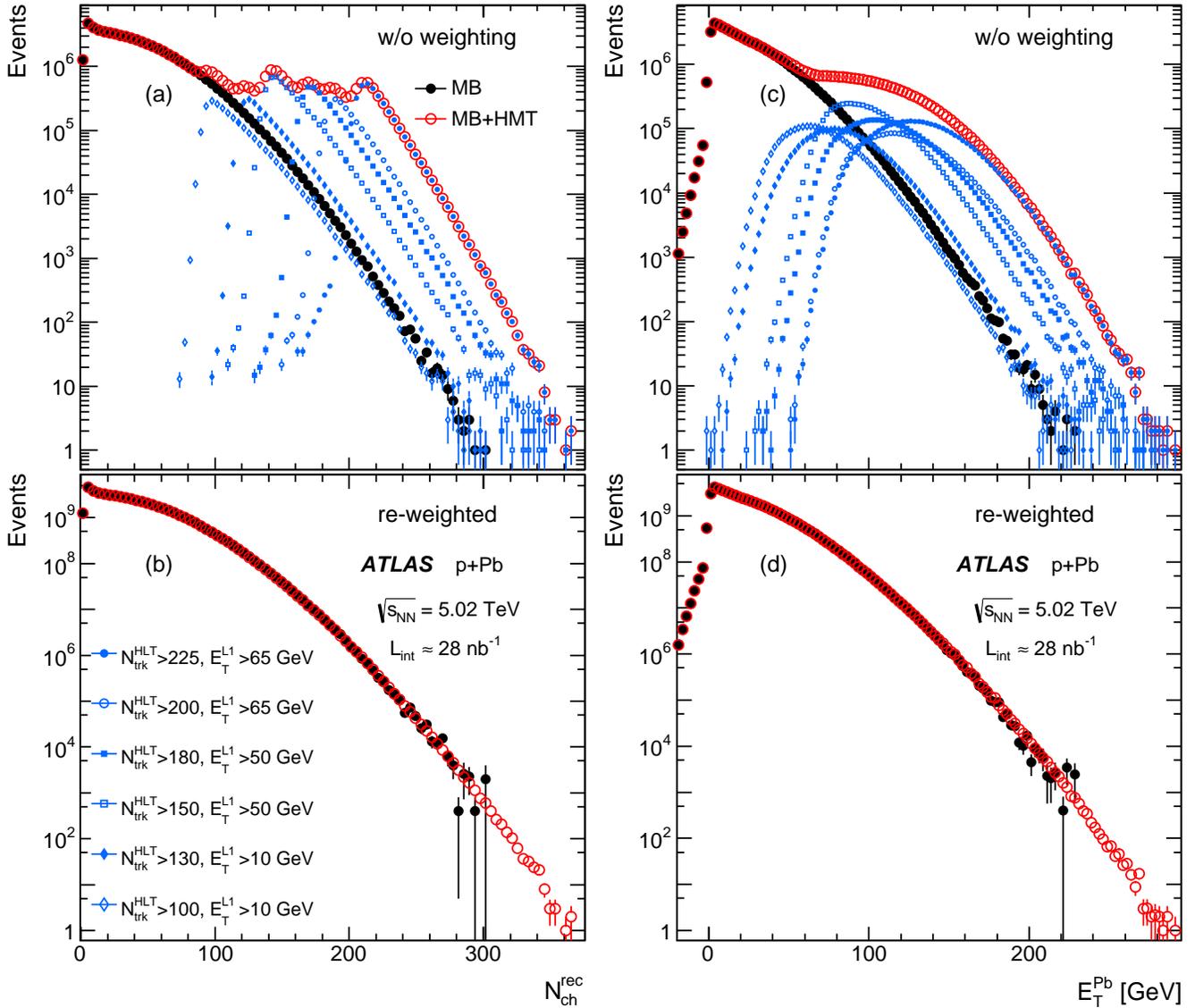}
\caption{\label{fig:2} The distributions of $\nchrec$ (left panels) and $\ETfcal$ (right panels) for MB and MB+HMT events before (top panels) and after (bottom panels) applying an event-by-event weight (see text). The smaller symbols in the top panels represent the distributions from the six HMT triggers listed in Table~\ref{tab:trig}.}
\end{figure}

Figure~\ref{fig:2} shows the distribution of $\nchrec$ (left panels) and $\ETfcal$ (right panels) for the MB and MB+HMT events before (top panels) and after (bottom panels) the re-weighting procedure. 
For MB-selected events, the re-weighted distribution differs from the original distribution by a constant factor, reflecting the average prescale. The multiple steps in the $\nchrec$ distribution (top-left panel) reflect the rapid turn-on behavior of individual HMT triggers in  $\nchrec$. The broad shoulder in the $\ETfcal$ distribution (top-right panel) is due to the finite width of the $\nchrec$ vs. $\ETfcal$ correlation, which smears the contributions from different HMT triggers in $\ETfcal$. All these structures disappear after the re-weighting procedure. The results of this analysis are obtained using the MB+HMT combined dataset, with event re-weighting. 

Due to the relatively slow turn-on of the HMT triggers as a function of $\ETfcal$ (Fig.~\ref{fig:2}(c)), the events selected in a given $\ETfcal$ range typically receive contributions from several HMT triggers with very different weights. Hence the effective increase in the number of events from the HMT triggers in the large $\ETfcal$ region is much smaller than the increase in the large $\nchrec$ region.

\begin{figure}[!h]
\centering
\includegraphics[width=0.5\columnwidth]{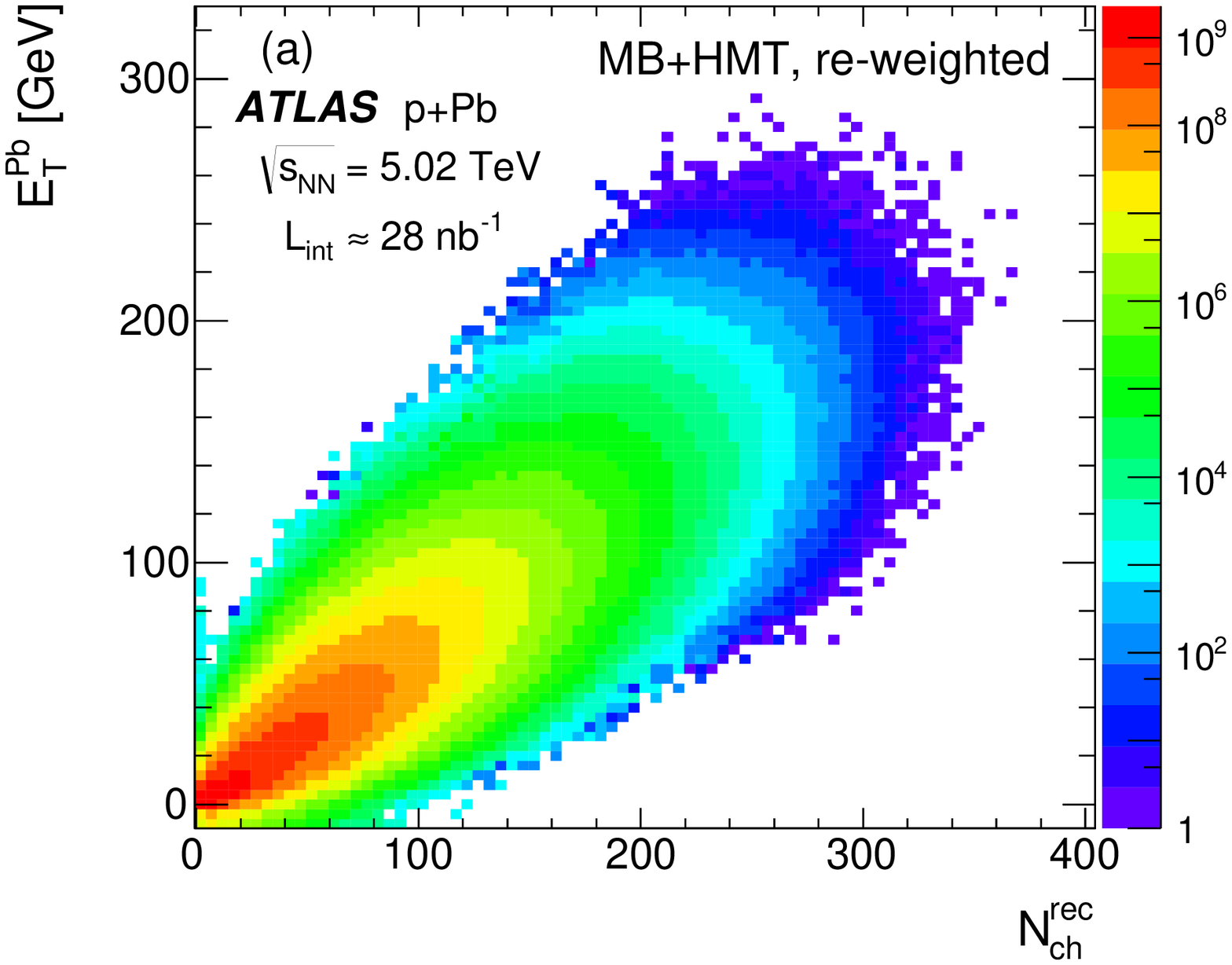}\includegraphics[width=0.5\columnwidth]{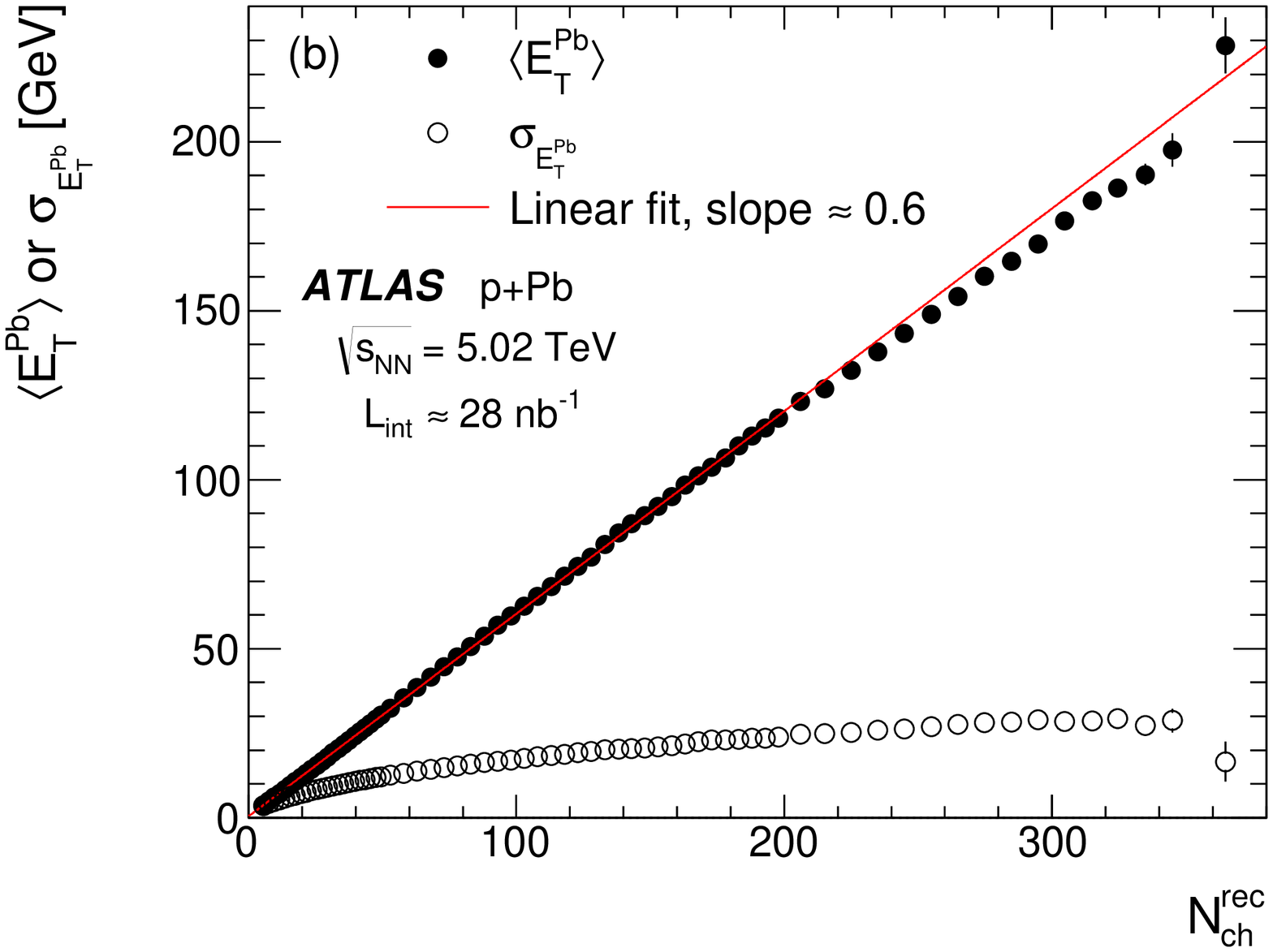}
\caption{\label{fig:2b} (a) Correlation between $\ETfcal$ and $\nchrec$ in MB+HMT events. (b) The mean $\left\langle\ETfcal\right\rangle$ and root-mean-square $\sigma_{\mathrm{E_{\rm T}^{\rm{Pb}}}}$ of the $\ETfcal$ distributions for slices of $\nchrec$. The line is a linear fit to all points.}
\end{figure}

Figure~\ref{fig:2b}(a) shows the correlation between $\ETfcal$ and $\nchrec$ from MB+HMT $\pPb$ events. This distribution is similar to that obtained for the MB events, except that the HMT triggers greatly extend the reach in both quantities. The $\ETfcal$ value grows with increasing $\nchrec$, suggesting that, on average, $\ETfcal$ in the nucleus direction correlates well with the particle production at mid-rapidity. On the other hand, the broad distribution of $\ETfcal$ at fixed $\nchrec$ also implies significant fluctuations. To study the relation between $\ETfcal$ and $\nchrec$, events are divided into narrow bins in $\nchrec$, and the mean and root-mean-square values of the $\ETfcal$ distribution are calculated for each bin. The results are shown in Fig.~\ref{fig:2b}(b). A nearly linear relation between $\left\langle\ETfcal\right\rangle$ and $\nchrec$ is observed. This relationship is used to match a given $\nchrec$ event class to the corresponding $\ETfcal$ event class. This approximately linear relation can also be parameterized (indicated by the solid line in Fig.~\ref{fig:2b}(b)) as
\begin{eqnarray} 
\label{eq:map}
\left\langle\ETfcal\right\rangle/\mathrm{GeV}\approx0.60\nchrec.
\end{eqnarray} 

The 2PC analysis is performed in different intervals of the event activity defined by either $\ETfcal$ or $\nchrec$. Table~\ref{tab:1} gives a list of representative event-activity classes, together with the fraction of MB+HMT events (after re-weighting as shown in Fig.~\ref{fig:2b}(a)) contained in each event class. The table also provides the average $\nchrec$ and $\ETfcal$ values for each event-activity class, as well as the efficiency-corrected number of charged particles within $|\eta|<2.5$ and $\pT>0.4$ GeV, $\nchave$. The event classes defined in narrow $\ETfcal$ or $\nchrec$ ranges are used for detailed studies of the centrality dependence of the 2PC, while the event classes in broad $\ETfcal$ or $\nchrec$ ranges are optimized for the studies of the $\pT$ dependence. As the number of events at large $\ETfcal$ is smaller than at large $\nchrec$, the main results in this paper are presented for event classes defined in $\nchrec$. 
\begin{table}[!h]
\centering
\small{
\begin{tabular}{|ccccc||ccccc|}\hline
\multicolumn{5}{|c||}{Event-activity classes based on $\nchrec$} & \multicolumn{5}{|c|}{Event-activity classes based on $\ETfcal$} \tabularnewline\hline
$\nchrec$ range & Fraction & $\ETfcalave$& $\nchrecave$ & $\nchave$    &$\ETfcal$ range& Fraction & $\ETfcalave$ & $\nchrecave$  & $\nchave$  \tabularnewline
                &          &  [GeV]      &              &            &   [GeV]       &          &  [GeV]       &               &  \tabularnewline\hline
 $<20$ &0.31   &  7.3&10.3&$12.6\pm0.6$&$<10$&0.28   &4.8&12.4&$ 15.4\pm0.7$\tabularnewline
 $[20, 40)$&0.27   &  18.6&29.1&$37.9\pm1.7$&$[10,23)$&0.26   &16.1&29.2&$ 38.1\pm1.7$\tabularnewline
 $[40, 60)$&0.19   &  30.8&48.8&$64.3\pm2.9$&$[23,37)$&0.19   &29.5&47.3&$ 62.3\pm2.8$\tabularnewline
 $[60, 80)$&0.12   &  42.8&68.6&$90.7\pm4.1$&$[37,52)$&0.12   &43.8&64.0&$ 84.7\pm3.8$\tabularnewline
 $[80,100)$&0.064   &  54.9&88.3&$117\pm5$&$[52,68)$&0.067   &58.8&80.4&$ 107\pm5$\tabularnewline
 $[100,120)$&0.029   &  66.4&108&$144\pm7$&$[68,83)$&0.028   &74.2&96.1&$ 128\pm6$\tabularnewline
 $[120,140)$&0.011   &  78.4&128&$170\pm8$&$[83,99)$&0.012      &89.7&111&$ 147\pm7$\tabularnewline
 $[140,160)$&0.0040   &  90.3&148&$196\pm9$&$[99,116)$&0.0043   &106&126&$ 168\pm8$\tabularnewline
 $[160,180)$&0.0013   &  102&168& $223\pm10$&$[116,132)$&0.0012   &122&141&$ 187\pm8$\tabularnewline
 $[180,200)$&$3.6\times10^{-4}$   &  113&187&$249\pm11$&$[132,148)$&$3.6\times10^{-4}$   &138&155&$ 206\pm9$\tabularnewline
 $[200,220)$&$9.4\times10^{-5}$   &  125&207&$276\pm12$&$[148,165)$&$1.0\times10^{-4}$   &155&169&$ 225\pm10$\tabularnewline
 $[220,240)$&$2.1\times10^{-5}$   &  134&227&$303\pm14$&$[165,182)$&$2.2\times10^{-5}$   &171&184&$ 244\pm11$\tabularnewline
 $[240,260)$&$4.6\times10^{-6}$   &  145&247&$329\pm15$&$[182,198)$&$4.6\times10^{-6}$   &188&196&$ 261\pm12$\tabularnewline
 $[260,290)$&$1.1\times10^{-6}$   &  157&269&$358\pm16$&$[198,223)$&$1.1\times10^{-6}$   &206&211&$ 281\pm13$\tabularnewline
 $[290,370)$&$8.9\times10^{-8}$   &  174&301&$393\pm18$&$[223,300)$&$9.6\times10^{-8}$   &232&230&$ 306\pm14$\tabularnewline\hline
 $<40$ &0.58   &  12.5&19.0&$24.4\pm1.1$&$<25$&0.59   &10.2&21.7&$ 28.0\pm1.3$\tabularnewline
 $[40, 80)$&0.32   &  35.3&56.4&$74.4\pm3.3$&$[25,50)$&0.27   &35.1&54.7&$ 72.2\pm3.3$\tabularnewline
 $[80,110)$&0.081   &  56.8&91.7&$122\pm6$&$[50,75)$&0.096   &61.5&81.4&$ 108\pm5$\tabularnewline
 $[110,140)$&0.023   &  74.2&121&$161\pm7$&$[75,100)$&0.025   &84.5&106&$ 141\pm6$\tabularnewline
 $[140,180)$&0.0053   &  93.0&153&$203\pm9$&$[100,130)$&0.0051   &110&130&$ 173\pm8$\tabularnewline
 $[180,220)$&$4.6\times10^{-4}$   &  116&192&$255\pm12$&$[130,165)$&$5.6\times10^{-4}$   &141&156&$ 208\pm9$\tabularnewline
 $[220,260)$&$2.6\times10^{-5}$   &  136&231&$307\pm14$&$[165,200)$&$2.7\times10^{-5}$   &174&186&$ 248\pm11$\tabularnewline
 $[260,370)$&$1.2\times10^{-6}$   &  158&271&$361\pm16$&$[200,300)$&$1.0\times10^{-6}$   &208&214&$ 284\pm13$\tabularnewline\hline
\end{tabular}}\normalsize
\caption{\label{tab:1} A list of event-activity classes defined in $\nchrec$ (left) and $\ETfcal$ (right) ranges, where the notation $[a,b)$ implies $a\leq\nchrec$~or~$\ETfcal<b$. For each event class, the fraction of MB+HMT events after trigger re-weighting (Fig.~\ref{fig:2b}(a)), the average values of $\ETfcalave$ and $\nchrecave$, and the efficiency-corrected average number of charged particles with $\pT>0.4$ GeV and $|\eta|<2.5$, $\nchave$, are also listed.}
\end{table}

\subsection{Two-particle correlation}
\label{sec:2PC}
For a given event class, the two-particle correlations are measured as a function of relative azimuthal angle, $\Dphi=\phi_a-\phi_b$, and relative pseudorapidity, $\Deta=\eta_a-\eta_b$, with $|\Deta|\leq \eta_{\Delta}^{\mathrm{max}}=5$. The labels $a$ and $b$ denote the two particles in the pair, which may be selected from different $\pT$ intervals. The particles $a$ and $b$ are conventionally referred to as the ``trigger'' and ``associated'' particles, respectively. The correlation strength, expressed in terms of the number of pairs per trigger particle, is defined as~\cite{Adare:2008ae,Alver:2009id,Aggarwal:2010rf}
\begin{eqnarray} 
\label{eq:ana1}
Y(\Dphi,\Deta) = \frac{\int B(\Dphi,\Deta) d\Dphi d\Deta }{\pi \eta_{\Delta}^{\mathrm{max}}}\left(\frac{S(\Dphi,\Deta)}{B(\Dphi,\Deta)}\right)\;,\;\; Y(\Dphi) = \frac{\int B(\Dphi) d\Dphi }{\pi} \left(\frac{S(\Dphi)}{B(\Dphi)}\right)\;,
\end{eqnarray}
where $S$ and $B$ represent pair distributions constructed from the same event and from ``mixed events''\cite{Adare:2008ae}, respectively, which are then normalized by the number of trigger particles in the event. These distributions are also referred to as per-trigger yield distributions. The mixed-event distribution, $B(\Dphi, \Deta)$, measures the distribution of uncorrelated pairs. The $B(\Dphi, \Deta)$ distribution is constructed by choosing the two particles in the pair from different events of similar $\nchrec$ (match to $|\Delta\nchrec|<10$ tracks), $\ETfcal$ (match to $|\Delta\ETfcal|<10$ GeV), and $\zvtx$ (match to $|\Delta\zvtx|<10$~mm), so that $B(\Dphi, \Deta)$ properly reflects the known detector effects in $S(\Dphi,\Deta)$. The one-dimensional (1-D) distributions $S(\Dphi)$ and $B(\Dphi)$ are obtained by integrating $S(\Dphi,\Deta)$ and $B(\Dphi,\Deta)$, respectively, over a $\Delta\eta$ range. The region $|\Delta\eta|<1$ is chosen to focus on the short-range features of the correlation functions, while the region $|\Delta\eta|>2$ is chosen to focus on the long-range features of the correlation functions. These two regions are hence referred to as the ``short-range region'' and ``long-range region'', respectively. The normalization factors in front of the $S/B$ ratio are chosen such that the $(\Dphi,\Deta)$-averaged value of $B(\Dphi,\Deta)$ and $\Dphi$-averaged value of $B(\Dphi)$ are both unity. When measuring $S$ and $B$, pairs are filled in one quadrant of the $(\Dphi, \Deta)$ space and then reflected to the other quadrants~\cite{Aad:2012gla}. To correct $S(\Dphi,\Deta)$ and $B(\Dphi,\Deta)$ for the individual inefficiencies of particles $a$ and $b$, the pairs are weighted by the inverse product of their tracking efficiencies $1/(\epsilon_a\epsilon_b)$. Remaining detector distortions not accounted for in the efficiency largely cancel in the $S/B$ ratio. 

\begin{figure}[!h]
\centering
\includegraphics[width=0.85\columnwidth]{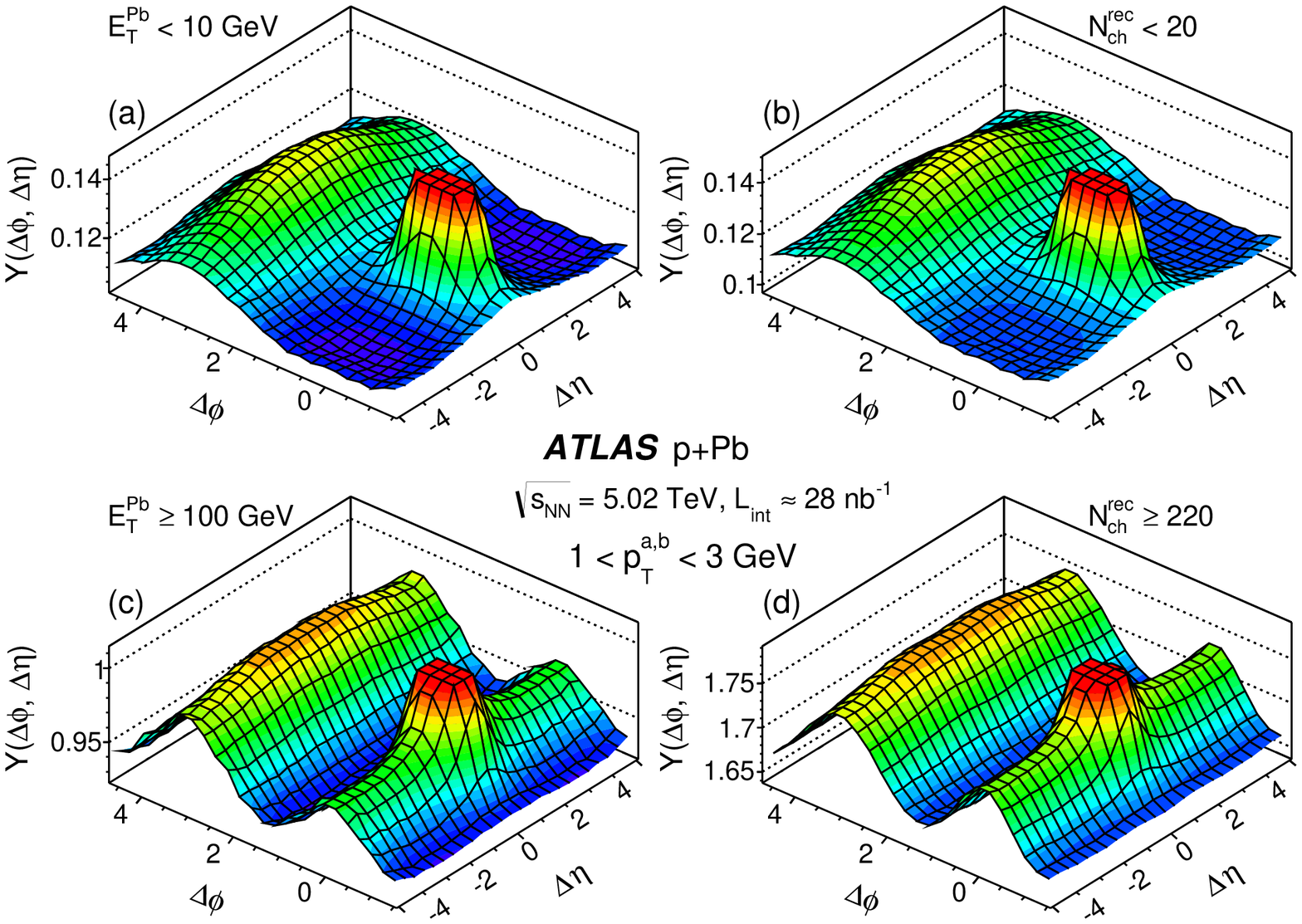}
\caption{\label{fig:method0} The 2-D correlation function in $\Dphi$ and $\Deta$ for the peripheral event class selected by either (a) $\ETfcal<10$ GeV or (b) $\nchrec<20$ and the central event class selected by either (c) $\ETfcal\geq100$ GeV or (d) $\nchrec\geq220$.}
\end{figure}

Examples of two-dimensional (2-D) correlation functions are shown in Fig.~\ref{fig:method0} for charged particles with $1<\ptab<3$~GeV in low-activity events, $\ETfcal<10$ GeV or $\nchrec<20$ in the top panels, and high-activity events, $\ETfcal>100$ GeV or $\nchrec>220$ in the bottom panels. The correlation for low-activity events shows a sharp peak centered at $(\Dphi,\Deta) = (0,0)$ due to short-range correlations for pairs resulting from jets, high-$\pT$ resonance decays, and Bose--Einstein correlations. The correlation function also shows a broad structure at $\Dphi\sim\pi$ from low-$\pT$ resonances, dijets, and momentum conservation that is collectively referred to as ``recoil''~\cite{Aad:2012gla} in the remainder of this paper. In the high-activity events, the correlation reveals a flat ridge-like structure at $\Dphi\sim0$ (the ``near-side'') that extends over the full measured $\Deta$ range. This $\Deta$ independence is quantified by integrating the 2-D correlation functions over $|\Dphi|<1$ to obtain $Y(\Deta) = \int_{|\Delta\phi|<1}Y(\Dphi,\Deta) \Dphi$. The yield associated with the near-side short-range correlation peak centered at $(\Dphi,\Deta) = (0,0)$ can then be estimated as
\begin{eqnarray} 
\label{eq:ana1b}
Y^{\mathrm{N-Peak}} = \int_{|\Delta\eta|<1} Y(\Deta) d\Deta-\frac{1}{5-\eta_{\Delta}^{\mathrm{min}}}\int_{\eta_{\Delta}^{\mathrm{min}}<|\Delta\eta|<5} Y(\Deta) d\Deta\;,
\end{eqnarray}
where the second term accounts for the contribution of uncorrelated pairs and the ridge component under the near-side peak. The default value of $Y^{\mathrm{N-Peak}}$ is obtained with a lower-end of the integration range of $\eta_{\Delta}^{\mathrm{min}}=2$, but the value of $\eta_{\Delta}^{\mathrm{min}}$ is varied from 2 to 4 to check the stability of $Y^{\mathrm{N-Peak}}$. The distribution at $\Dphi\sim\pi$ (the ``away-side'') is also broadened in high-activity events, consistent with the presence of a long-range component in addition to the recoil component~\cite{Aad:2012gla}. This recoil component can be estimated from the low-activity events and subtracted from the high-activity events using the procedure detailed in the next section.
 
\begin{figure}[!h]
\centering
\includegraphics[width=0.85\columnwidth]{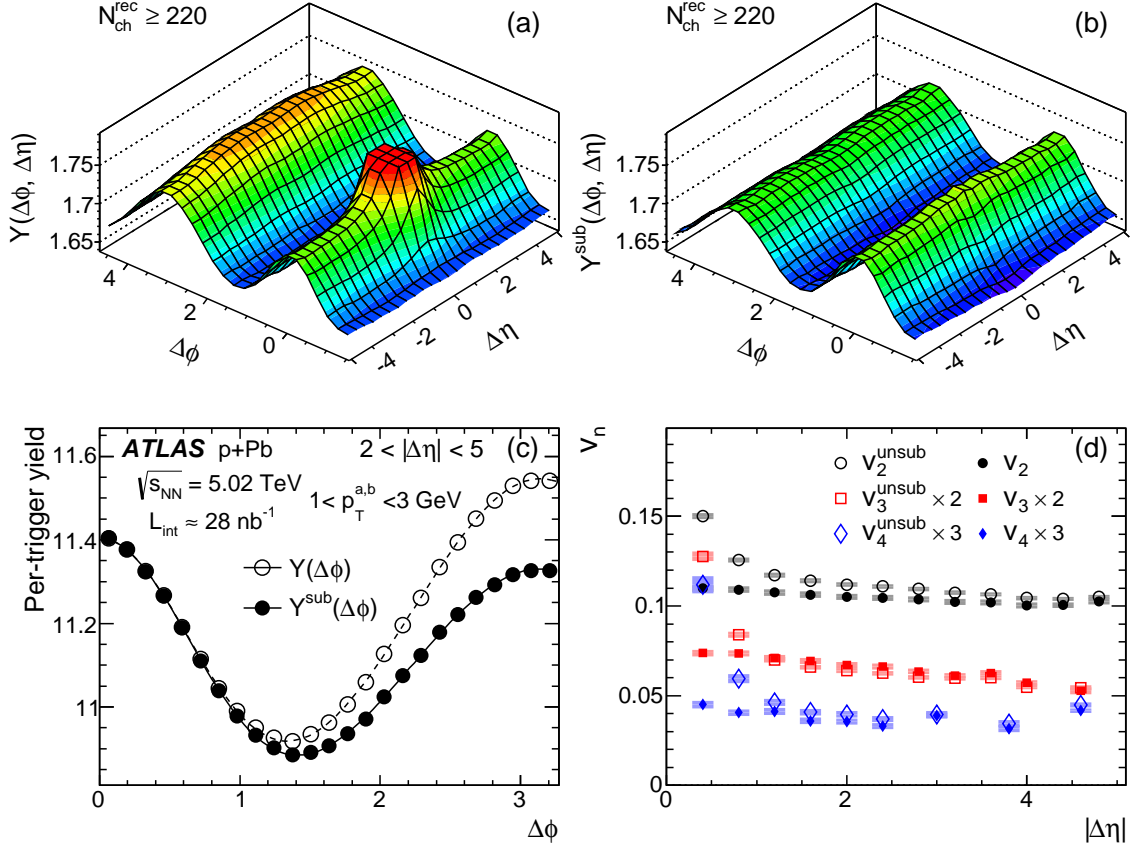}
\caption{\label{fig:method}  The 2-D correlation function in $\Dphi$ and $\Deta$ for events with $\nchrec\geq220$ (a) before and (b) after subtraction of the peripheral yield. Panel (c) shows the corresponding 1-D correlation functions in $\Dphi$ for pairs integrated over $2<|\Delta\eta|<5$ from panels (a) and (b), together with Fourier fits including the first five harmonics. Panel (d) shows the $2^{\mathrm{nd}}$,$3^{\mathrm{rd}}$, and $4^{\mathrm{th}}$-order Fourier coefficients as a function of $|\Deta|$ calculated from the 2-D distributions in panel (a) or panel (b), represented by the open or filled symbols, respectively. The error bars and shaded boxes are statistical and systematic uncertainties, respectively.}
\end{figure}

\subsection{Recoil subtraction}
\label{sec:recoil}
The correlated structure above a flat pedestal in the correlation functions is calculated using a zero-yield-at-minimum (ZYAM) method~\cite{Ajitanand:2005jj,Adare:2008ae} following previous measurements~\cite{CMS:2012qk,Abelev:2012ola,Aad:2012gla},
\begin{eqnarray} 
\label{eq:ana2}
Y^{\mathrm{corr}}(\Dphi,\Deta) = \frac{\int B(\Dphi,\Deta) d\Dphi d\Deta }{\pi \eta_{\Delta}^{\mathrm{max}}}\left(\frac{S(\Dphi,\Deta)}{B(\Dphi,\Deta)}-b_{_{\mathrm{ZYAM}}}\right),\; Y^{\mathrm{corr}}(\Dphi) = \frac{\int B(\Dphi) d\Dphi }{\pi} \left(\frac{S(\Dphi)}{B(\Dphi)}-b_{_{\mathrm{ZYAM}}}\right)\;,
\end{eqnarray}
where the parameter $b_{_{\mathrm{ZYAM}}}$ represents the pedestal formed by uncorrelated pairs. A second-order polynomial fit to the 1-D $Y(\Delta \phi)$ distribution in the long-range region is used to find the location of the minimum point, $\Dphi_{_{\mathrm{ZYAM}}}$, and from this the value of $b_{_{\mathrm{ZYAM}}}$ is determined and subtracted from the 2-D correlation function. The $Y^{\mathrm{corr}}(\Dphi,\Deta)$ functions differ, therefore, by a constant from the $Y(\Dphi,\Deta)$ functions, such as those in Fig.~\ref{fig:method0}. 

In low-activity events, $Y^{\mathrm{corr}}(\Dphi,\Deta)$ contains mainly the short-range correlation component and the recoil component. In high-activity events, the contribution from the long-range ``ridge'' correlation also becomes important. This long-range component of the correlation function in a given event class is obtained by estimating the short-range correlation component using the peripheral events and is then subtracted,
\begin{eqnarray} 
\label{eq:ana3}
Y^{\mathrm{sub}}(\Dphi,\Deta) = Y(\Dphi,\Deta)-\alpha Y_{\mathrm{peri}}^{\mathrm{corr}}(\Dphi,\Deta), \; \; Y^{\mathrm{sub}}(\Dphi) = Y(\Dphi)-\alpha Y_{\mathrm{peri}}^{\mathrm{corr}}(\Dphi),
\end{eqnarray}
where the $Y^{\mathrm{corr}}$ in a low-activity or peripheral event class, denoted by $Y_{\mathrm{peri}}^{\mathrm{corr}}$, is used to estimate and subtract (hence the superscript ``sub'' in Eq.~(\ref{eq:ana3})) the short-range correlation at the near-side and the recoil at the away-side. The parameter $\alpha$ is chosen to adjust the near-side short-range correlation yield in the peripheral events to match that in the given event class for each $\pta$ and $\ptb$ combination,  $\alpha=Y^{\mathrm{N-Peak}}/Y^{\mathrm{N-Peak}}_{\mathrm{peri}}$. This scaling procedure is necessary to account for enhanced short-range correlations and away-side recoil in higher-activity events, under the assumption that the relative contribution of the near-side short-range correlation and away-side recoil is independent of the event activity. A similar rescaling procedure has also been used by the CMS Collaboration~\cite{Chatrchyan:2013nka}. The default peripheral event class is chosen to be $\ETfcal<\eT^0=10$ GeV. However, the results have also been checked with other $\eT^0$ values, as well as with a peripheral event class defined by $\nchrec<20$. In the events with the highest multiplicity, the value of $\alpha$ determined with the default peripheral event class varies from $\sim 2$ at $\pT\approx 0.5$ GeV to $\sim 1$ for $\pT>3$ GeV, with a $\pT$-dependent uncertainty of 3--5\%. 

The uncertainty on $b_{_{\mathrm{ZYAM}}}$ only affects the recoil-subtracted correlation functions through the $Y_{\mathrm{peri}}^{\mathrm{corr}}$ term in Eq.~(\ref{eq:ana3}). This uncertainty is usually very small in high-activity $\pPb$ collisions, due to their much larger pedestal level than for the peripheral event class.

Figures~\ref{fig:method}(a) and (b) show, respectively, the 2-D correlation functions before and after the subtraction procedure given by Eq.~(\ref{eq:ana3}). Most of the short-range peak and away-side recoil structures are removed by the subtraction, and the remaining distributions exhibit a $\Dphi$-symmetric double-ridge that is almost independent of $\Deta$. Figure~\ref{fig:method}(c) shows the corresponding 1-D correlation functions before and after recoil subtraction in the long-range region of $|\Deta|>2$. The distribution at the near-side is not affected since the near-side short-range peak is narrow in $\eta$ (Fig.~\ref{fig:method}(a)), while the away-side distribution is reduced due to the removal of the recoil component.

\subsection{Extraction of the azimuthal harmonics associated with long-range correlation}
\label{sec:fourier}
The azimuthal structure of the long-range correlation is studied via a Fourier decomposition similar to the approach used in the analysis of Pb+Pb collisions~\cite{Aamodt:2011by,Aad:2012bu},
\begin{eqnarray}
\label{eq:four1}
 Y^{\mathrm{sub}}(\Dphi)=\frac{\int Y^{\mathrm{sub}}(\Dphi)d\Dphi}{\pi}\left(1 + \sum_n 2v_{n,n}\cos(n\Delta\phi) \right),
\end{eqnarray}
where $v_{n,n}$ are the Fourier coefficients calculated via a discrete Fourier transformation,
\begin{eqnarray} 
v_{n,n}= \frac{\sum_{m=1}^{N} \cos (n\Delta\phi_m) Y^{\mathrm{sub}}(\Dphi_m)}{\sum_{m=1}^N Y^{\mathrm{sub}}(\Dphi_m)}\;,
\end{eqnarray}
where $N=24$ is the number of $\Dphi$ bins from 0 to $\pi$. The first five Fourier coefficients are calculated as a function of $\pta$ and $\ptb$ for each event-activity class.

The azimuthal anisotropy coefficients for single particles, $v_n$, can be obtained via the factorization relation commonly used for heavy-ion collisions~\cite{Aamodt:2011by,Aad:2012bu,Adcox:2002ms},
\begin{eqnarray} 
\label{eq:four2}
v_{n,n}(\pta,\ptb)=v_n(\pta)v_n(\ptb).
\end{eqnarray}
From this the $\pT$ dependence of $v_n$ for $n=2$--5 are calculated as
\begin{eqnarray} 
\label{eq:four3}
v_n(\pta) = v_{n,n}(\pta,\ptref)/\sqrt{v_{n,n}(\ptref,\ptref)}\;,
\end{eqnarray}
where the default transverse momentum range for the associated particle (b) is chosen to be $1<\ptref<3$ GeV, and the Fourier coefficient as a function of the transverse momentum of the trigger particle is denoted by $v_n(\pta)$, or simply $v_n(\pT)$ where appropriate. The extraction of $v_1$ requires a slight modification and is discussed separately in Sec.~\ref{sec:v1}. The factorization behavior is checked by comparing the $v_n(\pta)$ obtained for different $\ptref$ ranges, as discussed in Sec.~\ref{sec:result2}.

A similar Fourier decomposition procedure is also carried out for correlation functions without peripheral subtraction, i.e. $Y(\Dphi)$. The harmonics obtained in this way are denoted by $v_{n,n}^{\mathrm{unsub}}$ and $v_{n}^{\mathrm{unsub}}$, respectively.

Figure~\ref{fig:method}(d) shows the azimuthal harmonics obtained by Fourier decomposition of the $Y(\Dphi,\Deta)$ and $Y^{\mathrm{sub}}(\Dphi,\Deta)$ distributions in Figs.~\ref{fig:method}(a) and (b) for different, narrow slices of $\Deta$. The resulting $v_{n}^{\mathrm{unsub}}$ and $v_{n}$ values are plotted as a function of $\Deta$ for $n=2$, 3 and 4. The $v_n$ values are much smaller than $v_n^{\mathrm{unsub}}$ for $|\Deta|<1$, reflecting the removal of the short-range correlations at the near-side. The $v_2$ values are also systematically smaller than $v_2^{\mathrm{unsub}}$ for $|\Deta|>1$, reflecting the removal of the away-side recoil contribution. 

\subsection{Systematic uncertainties}
\label{sec:sys}
The systematic uncertainties in this analysis arise from pair acceptance, the ZYAM procedure, tracking efficiency, Monte Carlo consistency, residual pileup, and the recoil subtraction. Each source is discussed separately below.

The correlation functions rely on the pair acceptance functions, $B(\Delta\phi, \Delta\eta)$ and $B(\Delta\phi)$ in Eq.~(\ref{eq:ana1}), to reproduce detector acceptance effects in the signal distribution. A natural way of quantifying the influence of detector effects on $v_{n,n}$ and $v_n$ is to express the single-particle and pair acceptance functions as Fourier series, similar to Eq.~(\ref{eq:four1}). The resulting coefficients for pair acceptance $v_{n,n}^{\mathrm{det}}$ are the product of those for the two single-particle acceptances $v_{n}^{\mathrm{det,a}}$ and $v_{n}^{\mathrm{det,b}}$. In general, the pair acceptance function in $\Dphi$ is quite flat: the maximum fractional variation from its average value is observed to be less than 0.001 for pairs integrated in 2 $< |\Delta\eta| <$ 5, and the corresponding $|v_{n,n}^{\mathrm{det}}|$ values are found to be less than $2\times 10^{-4}$. These $v_{n,n}^{\mathrm{det}}$ values are expected to mostly cancel in the correlation function, and only a small fraction contributes to the uncertainties in the pair acceptance function. Possible residual effects on the pair acceptance are evaluated following Ref.~\cite{Aad:2012bu}, by varying the criteria for matching in $\nchrec$, $\ETfcal$, and $\zvtx$.  In each case, the residual $v_{n,n}^{\mathrm{det}}$ values are evaluated by a Fourier expansion of the ratio of the pair acceptances before and after the variation. This uncertainty varies in the range of (5--8)$\times$ $10^{-6}$. It is negligible for $v_2$ and $v_3$, but become sizable for higher-order harmonics, particularly at low $\pT$, where the $v_n$ values are small. 

 As discussed in Sec.~\ref{sec:recoil}, the value of $b_{_{\mathrm{ZYAM}}}$ is determined by a second-order polynomial fit of the $Y(\Delta \phi)$ distribution. The stability of the fit is studied by varying the $\Dphi$ range in the fit. The uncertainty in $b_{_{\mathrm{ZYAM}}}$ depends on the local curvature around $\Dphi_{_{\mathrm{ZYAM}}}$, and is estimated to be 0.0003--0.001 of the minimum value of $Y(\Delta \phi)$. This uncertainty contributes directly to $Y^{\mathrm{corr}}(\Dphi)$, but contributes to $Y^{\mathrm{sub}}(\Dphi)$ and $v_n$ indirectly through the peripheral subtraction (see Eq.~(\ref{eq:ana3})). The resulting uncertainty on $v_n$ is found to be less than $2\%$, for all $n$.

The values of per-trigger yields, $Y(\Dphi)$, $Y^{\mathrm{corr}}(\Dphi)$, and $Y^{\mathrm{sub}}(\Dphi)$, are sensitive to the uncertainty on the tracking efficiency correction for the associated particles. This uncertainty is estimated by varying the track quality cuts and the detector material in the simulation, re-analyzing the data using corresponding Monte Carlo efficiencies and evaluating the change in the extracted yields. The resulting uncertainty is estimated to be 2.5\% due to the track selection and 2--3\% related to our limited knowledge of the detector material. The $v_{n,n}$ and $v_n$ values depend only on the shape of the $Y^{\mathrm{sub}}(\Dphi)$ distribution and hence are not sensitive to the tracking efficiency. 

The analysis procedure is also validated by measuring $v_n$ values in fully simulated HIJING events~\cite{Agostinelli:2002hh,Aad:2010ah} and comparing them to those measured using the generated particles. A small but systematic difference between the two results are included in the systematic uncertainties.

Nearly all of the events containing pileup are removed by the procedure described in Sec.~\ref{sec:sel}. The influence of the residual pileup is evaluated by relaxing the pileup rejection criteria and then calculating the change in the per-trigger yields and $v_n$ values. The differences are taken as an estimate of the uncertainty, and are found to be negligible in low event-activity classes, and increase to 2\% for events with $\ETfcal>200$ GeV or $\nchrec>300$.

According to Table~\ref{tab:1}, the low-activity events used in the peripheral subtraction ($\ETfcal<\eT^0=10$ GeV) correspond to 28\% of the MB-triggered events. The pair distributions for these events may contain a small genuine long-range component, leading to a reduction of the long-range correlation signal in a high-activity class via the peripheral subtraction procedure. The influence of this over-subtraction is evaluated by varying the definition of the low-activity events in the range of $\eT^0=5$ GeV to $\eT^0=20$~GeV. The $Y^{\mathrm{sub}}(\Dphi)$ and $v_n$ values are calculated for each variation. The $v_n$ values are found to decrease approximately linearly with increasing $\eT^0$. The amount of over-subtraction can be estimated by extrapolating $\eT^0$ to zero. The estimated changes of $v_n$ and $Y^{\mathrm{sub}}(\Dphi)$ vary from less than 1\% for $\ETfcal>100$ GeV or $\nchrec>150$, and increase for lower event-activity classes approximately as $1.5/\nchrec$. The relative change of $v_n$ is also found to be independent of $\pT$. As a cross-check, the analysis is repeated by defining peripheral events as $\nchrec <$ 20. The variation of $v_n$ values is found to be consistent with the variation from varying $\eT^0$.

The stability of the scale factor, $\alpha$, is evaluated by varying the $\Delta\eta$ window of the long-range region in Eq.~(\ref{eq:ana1b}). A 3--5\% uncertainty is quoted for $\alpha$ from these variations. The resulting uncertainty on $v_n$ for $n=2$--5 is within 1\% at low $\pT$ ($<$ 4 GeV), and increases to $\sim$10\% at the highest $\pT$. However, the $v_1$ extraction is directly affected by the subtraction of the recoil component, and hence the $v_1$ value is very sensitive to the uncertainty in $\alpha$. The estimated uncertainty is 8--12\% for $\pT<1$ GeV and is about 20--30\% for $\pT>3$ GeV.

The different sources of the systematic uncertainties described above are added in quadrature to give the total systematic uncertainties for per-trigger yields and $v_n$, which are summarized in Tables~\ref{tab:tabpty} and \ref{tab:tabvn}, respectively. The systematic uncertainty quoted for each source usually covers the maxmium uncertainty over the measured $\pT$ range and event-activity range. However, since $v_1(\pT)$ changes sign within 1.5--2.0 GeV (see Fig.~\ref{fig:v1b}), the relative uncertainties are quoted for $\pT<1$ GeV and $\pT>3$ GeV. The uncertainty of pair acceptance, which is less than $8\times 10^{-6}$ for $v_{n,n}$, was converted to percent uncertainties. This uncertainty can be significant at high $\pT$.
\begin{table}[!h]
\centering
\begin{tabular}{|l|c|}\hline
Residual pair acceptance [$\%$] & 1--2 \tabularnewline\hline
ZYAM procedure [$\%$] & 0.2--1.5 \tabularnewline\hline
Tracking efficiency \& material [$\%$]   & 4.2  \tabularnewline\hline
Residual pileup [$\%$]     & 0--2 \tabularnewline\hline
\end{tabular}
\caption{\label{tab:tabpty} Summary of relative systematic uncertainties for $Y(\Dphi)$, $Y^{\mathrm{corr}}(\Dphi)$ and $Y^{\mathrm{sub}}(\Dphi)$.}
\end{table}

\begin{table}[!h]
\centering
\begin{tabular}{|l|c|c|c|c|c|}\hline
  & $n=1$ &$n=2$ & $n=3$ & $n=4$ & $n=5$ \tabularnewline\hline
Residual pair acceptance [\%] & 1.0--5.0 & $<$0.5 & 1.0--4.0 & 7.0--12 & 7.0--20\tabularnewline\hline
ZYAM procedure [\%] & 0.6 & 0.3 & 0.3 & 0.5 & 0.6 \tabularnewline\hline
Tracking efficiency\& material [\%] & 1.0 & 0.4 & 0.8 & 1.2 & 2.4  \tabularnewline\hline
Monte Carlo consistency [\%] & 4.0& 1.0 & 2.0 & 4.0 & 8.0 \tabularnewline\hline
Residual pileup [\%]   & 0--2.0  & 0--2.0 & 0--2.0 & 0--2.0 & 0--2.0\tabularnewline\hline
Uncertainty on scale factor $\alpha$ [\%] & 8.0--30 &  0.2--10 & 0.2--12 & 0.2--14 & 1.0--14 \tabularnewline\hline
\shortstack{Choice of peripheral events \\for $\nchrec >$ 160 or $\ETfcal>$100 GeV [\%]} & 4.0 & 1.0 & 1.0 & 2.0 &  4.0 \tabularnewline\hline
\end{tabular}
\caption{\label{tab:tabvn} Summary of relative systematic uncertainties on $v_n$, for $n=1$ to 5.}
\end{table}

\section{Results}
\label{sec:result}
\subsection{Correlation functions and integrated yields}
\label{sec:result1}
\begin{figure}[!h]
\centering
\includegraphics[width=1\columnwidth]{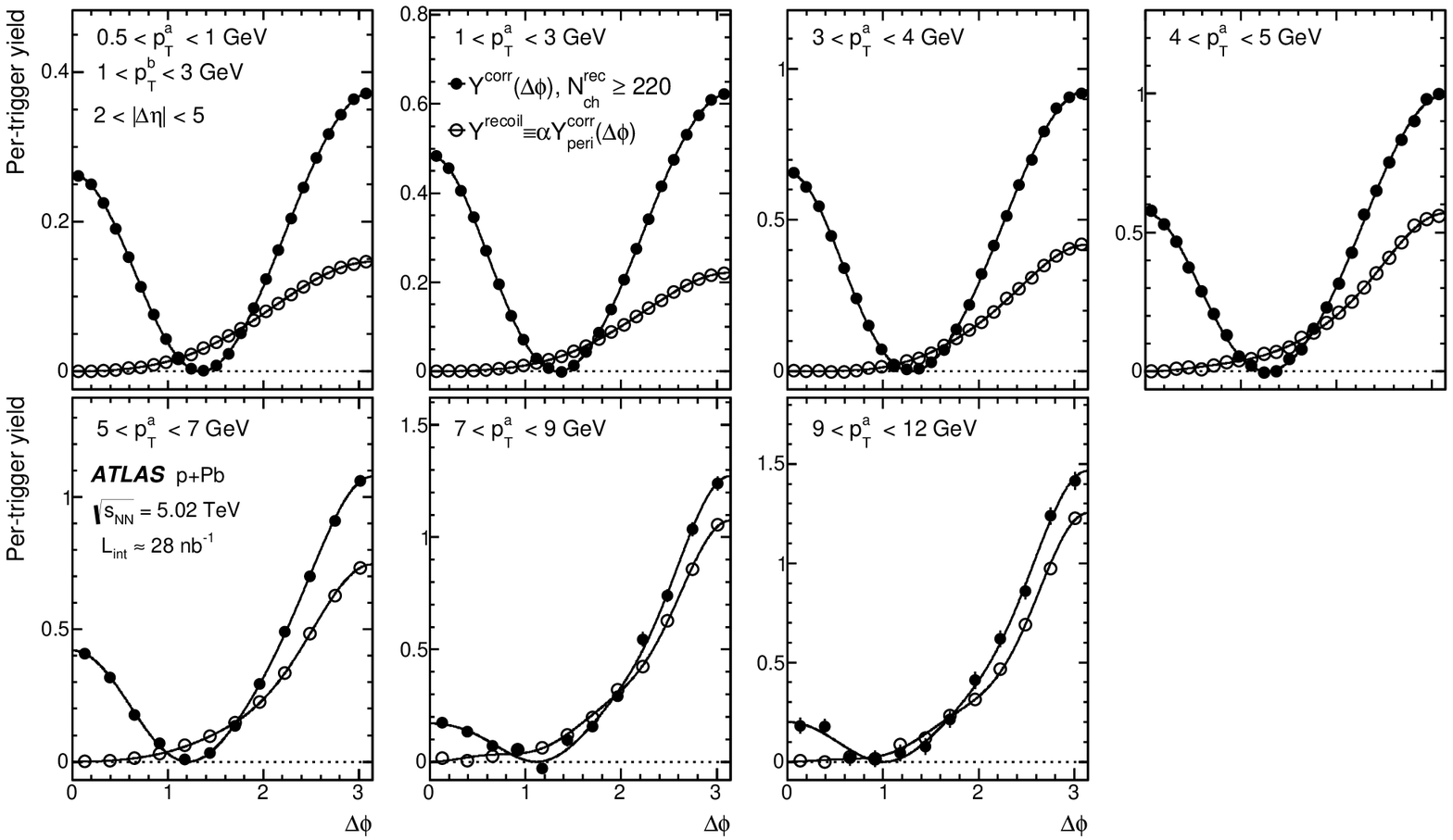}
\caption{\label{fig:res1} The per-trigger yield distributions $Y^{\mathrm{corr}}(\Dphi)$ and $Y^{\mathrm{recoil}}(\Dphi)$ for events with $\nchrec\geq220$ in the long-range region \mbox{$|\Delta\eta|>2$}. The distributions are shown for $1<\ptb<3$ GeV in various $\pta$ ranges. They are compared to the recoil contribution estimated from a peripheral event class defined by $\ETfcal<10$ GeV using a rescaling procedure (see Eq.~(\ref{eq:ana3}) and discussion around it). The curves are Fourier fits including the first five harmonics.}
\end{figure}

Figure~\ref{fig:res1} shows the 1-D correlation functions after the ZYAM procedure, $Y^{\mathrm{corr}}(\Dphi)$, in various ranges of $\pta$ for a fixed $\ptb$ range of 1--3 GeV. The correlation functions are obtained in the long-range region ($|\Delta\eta|>2$) and are shown for events selected by $\nchrec\geq220$. This event class contains a small fraction ($3\times10^{-5}$) of the minimum-bias $\pPb$ events with highest multiplicity. The correlation functions are compared to the distributions of the recoil component, $\alpha Y^{\mathrm{corr}}_{\mathrm{peri}}(\Dphi)$ in Eq.~(\ref{eq:ana3}), estimated from the peripheral event class defined by $\ETfcal<10$ GeV. The scale factor $\alpha$ is chosen such that the near-side short-range yield matches between the two event classes (see Eq.~(\ref{eq:ana3}) and discussion around it). Figure~\ref{fig:res1} shows a clear near-side excess in the full $\pta$ range studied in this analysis. An excess above the estimated recoil contribution is also observed on the away-side over the same $\pT$ range.
 
\begin{figure}[!h]
\centering
\includegraphics[width=0.9\columnwidth]{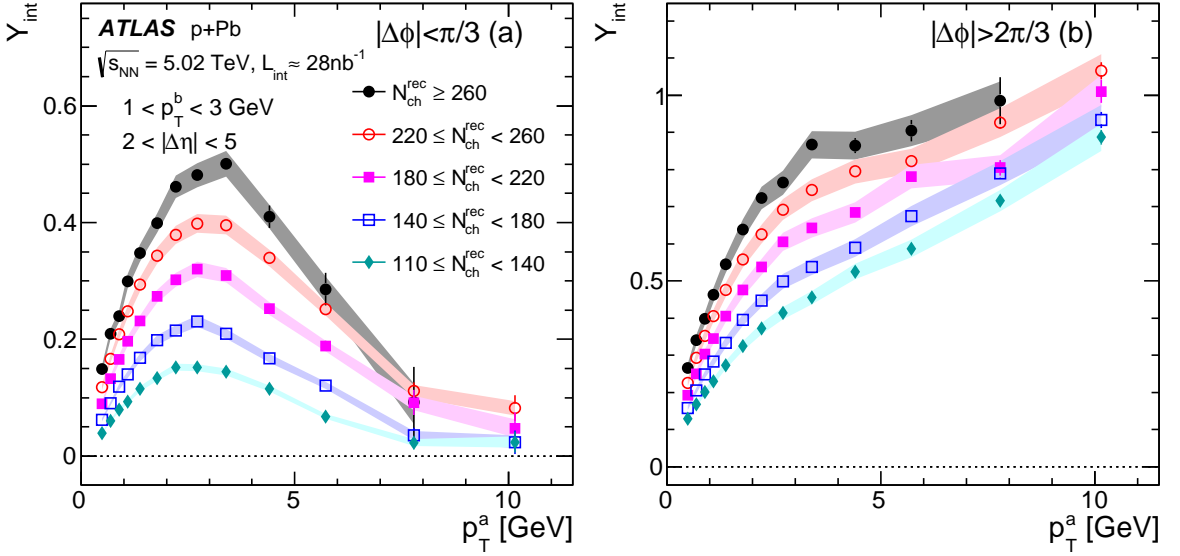}
\caption{\label{fig:res2} Integrated per-trigger yields $\Yint$ as a function of $\pta$ for $1<\ptb<3$~GeV, for events in various $\nchrec$ ranges on (a) the near-side and (b) the away-side. The errors bars and shaded bands represent the statistical and systematic uncertainties, respectively.}
\end{figure}

To further quantify the properties of the long-range components, the $Y^{\mathrm{corr}}(\Dphi)$ distributions are integrated over $|\Dphi|<\pi/3$ and $|\Dphi|>2\pi/3$, similar to the procedure used in previous analyses~\cite{Aad:2012gla,Abelev:2012ola}. The integrated yields, $\Yint$, are obtained in several event classes and are plotted as a function of $\pta$ in Fig.~\ref{fig:res2}. The near-side yields increase with trigger $\pT$, reach a maximum at $\pT\sim3$ GeV, and then decrease to a value close to zero at $\pT>10$ GeV. This trend is characteristic of the $\pT$ dependence of the Fourier harmonics in A+A collisions. In contrast, the away-side yields show a continuous increase across the full $\pT$ range, due to the contribution of the recoil component that mostly results from dijets.

\begin{figure}[!h]
\centering
\includegraphics[width=0.9\columnwidth]{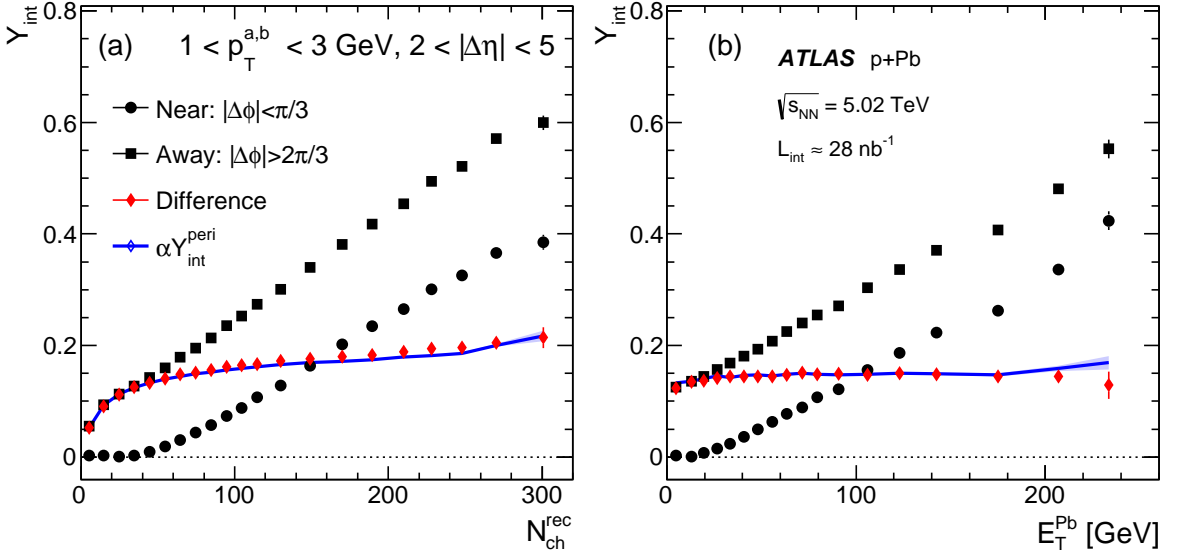}
\caption{\label{fig:res3} The integrated per-trigger yield, $\Yint$, on the near-side (circles), the away-side (squares) and their difference (diamonds) as a function of (a) $\nchrec$ and (b) $\ETfcal$ for pairs in $2<|\Delta\eta|<5$ and  $1<\ptab<3$ GeV. The yield difference is compared to the estimated recoil contribution in the away-side (solid lines). The error bars or the shaded bands represent the combined statistical and systematic uncertainties.}
\end{figure}

Figure~\ref{fig:res3} shows the centrality dependence of the long-range integrated yields for the event-activity based on $\nchrec$ (left) and $\ETfcal$ (right) for particles in $1<\ptab<3$ GeV range. The near-side yield is close to zero in low-activity events and increases with $\ETfcal$ or $\nchrec$. The away-side yield shows a similar increase as a function of $\ETfcal$ or $\nchrec$, but it starts at a value significantly above zero. The yield difference between these two regions is found to vary slowly with $\ETfcal$ or $\nchrec$, indicating that the growth in the integrated yield with increasing event activity is similar on the near-side and the away-side. This behavior suggests the existence of an away-side long-range component that has a magnitude similar to the near-side long-range component.

Figure~\ref{fig:res3} also shows (solid lines) the recoil component estimated from the low event-activity class ($\ETfcal<10$ GeV) via the rescaling procedure discussed in Sec.~\ref{sec:recoil}. The yield difference between the away-side and the near-side in this $\pT$ range is reproduced by this estimate of the recoil component. In other $\pT$ ranges, a systematic difference between the recoil component and the yield difference is observed and is attributed to the contribution of a genuine dipolar flow, $v_{1,1}$, to the correlation function (see discussion in Sec.~\ref{sec:v1}).

To quantify the $\Dphi$ dependence of the measured long-range correlations, the first five harmonics of the correlation functions, $v_1$ to $v_5$, are extracted via the procedure described in Sec.~\ref{sec:fourier}. The following section summarizes the results for $v_2$--$v_5$, and the results for $v_1$ are discussed in Sec.~\ref{sec:v1}.
 
\subsection{Fourier coefficients $v_2$--$v_5$}
\label{sec:result2}
\begin{figure}[!h]
\centering
\includegraphics[width=1\columnwidth]{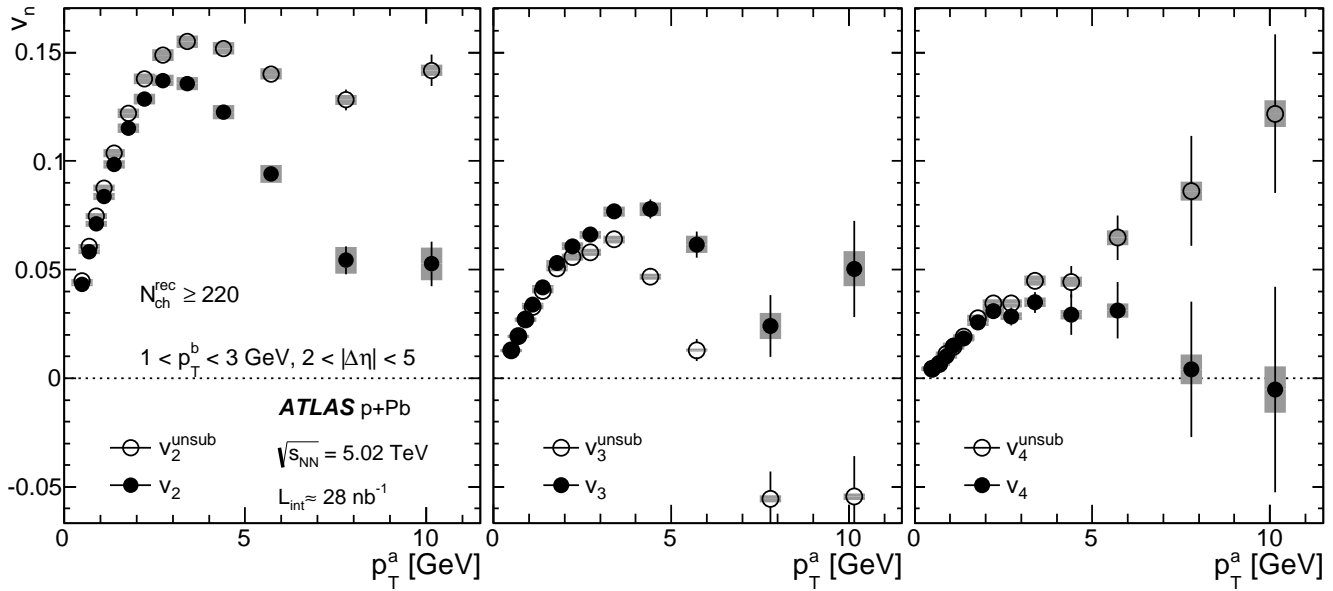}
\caption{\label{fig:resb1} The Fourier coefficients $v_2$, $v_3$, and $v_4$ as a function of $\pta$ extracted from the correlation functions for events with $\nchrec\geq220$, before (denoted by $v_n^{\rm{unsub}}$) and after (denoted by $v_n$) the subtraction of the recoil component. Each panel shows the results for one harmonic. The pairs are formed from charged particles with $1<\ptref<3$ GeV and $|\Delta\eta|>2$. The error bars and shaded boxes represent the statistical and systematic uncertainties, respectively.}
\end{figure}

Figure~\ref{fig:resb1} shows the $v_2$, $v_3$, and $v_4$ obtained
using the 2PC method described in Sec.~\ref{sec:fourier} for \mbox{$1<\ptref<3$~GeV}. The results are shown both before (denoted by $v_n^{\rm{unsub}}$) and after the subtraction of the recoil component (Eq.~(\ref{eq:ana3})). The recoil contribution affects slightly the $v_n$ values for trigger $\pt<3$ GeV, but becomes increasingly important for higher trigger $\pT$ and higher-order harmonics. This behavior is expected as the dijet contributions, the dominant contribution to the recoil component, increase rapidly with $\pt$ (for example see Fig.~\ref{fig:res1} or  Ref.~\cite{Aad:2012bu}).  At high $\pT$, the contribution of dijets appears as a narrow peak at the away-side, leading to $v_n^{\rm{unsub}}$ coefficients with alternating sign: $(-1)^n$~\cite{Aad:2012bu}. In contrast, the $v_n$ values after recoil subtraction are positive across the full measured $\pt$ range. Hence, the recoil subtraction is necessary for the reliable extraction of the long-range correlations, especially at high $\pt$. 

\begin{figure}[!h]
\centering
\includegraphics[width=1\columnwidth]{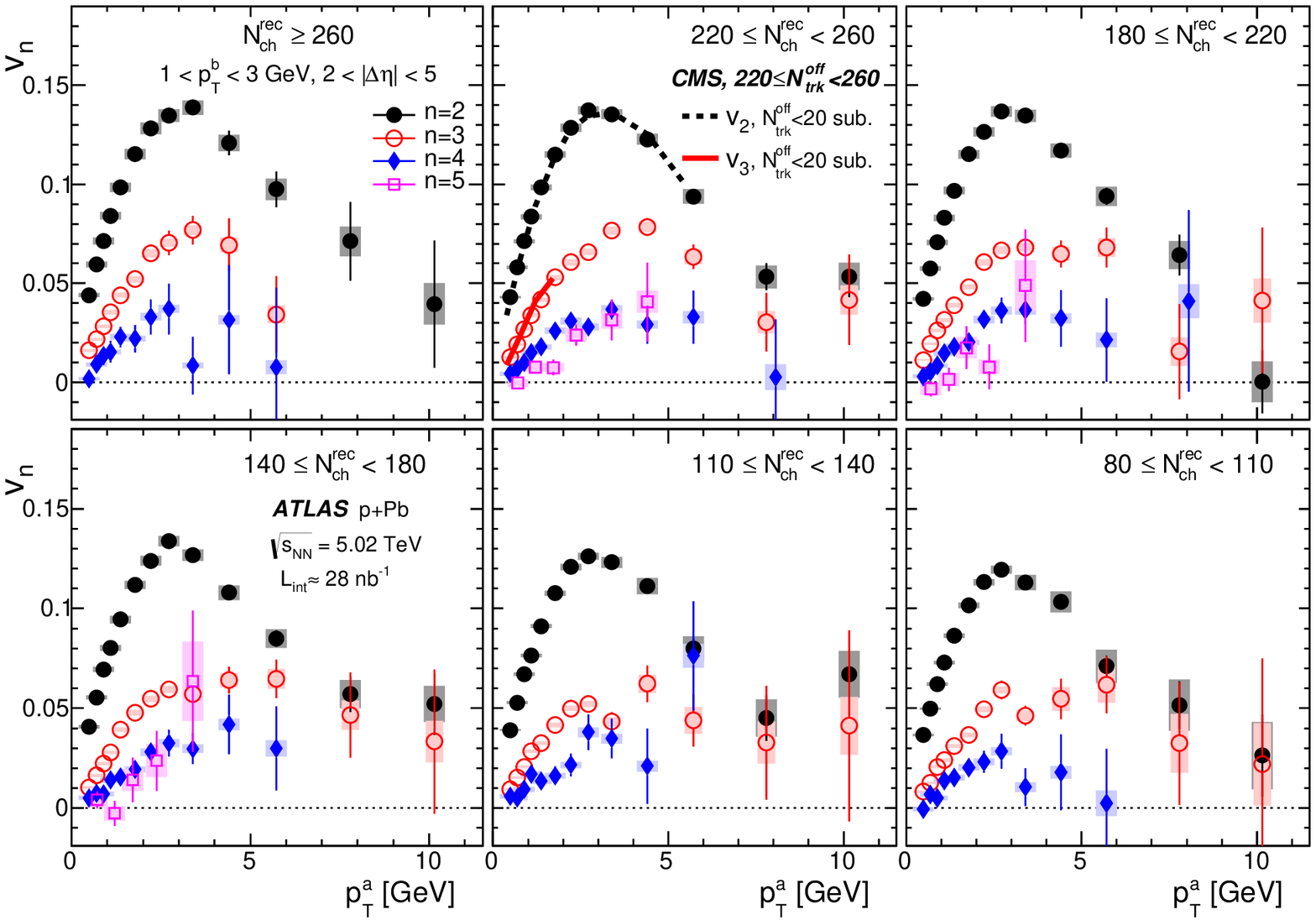}
\caption{\label{fig:resb1b} The $v_n(\pta)$ with $n=2$ to 5 for six $\nchrec$ event-activity classes obtained for $|\Deta|>2$ and the $\ptref$ range of 1--3 GeV. The error bars and shaded boxes represent the statistical and systematic uncertainties, respectively. Results in $220\leq\nchrec<260$ are compared to the CMS data~\cite{Chatrchyan:2013nka} obtained by subtracting the peripheral events (the number of offline tracks $N_{\mathrm{trk}}^{\mathrm{off}}<20$), shown by the solid and dashed lines.}
\end{figure}

Figure~\ref{fig:resb1b} shows the trigger $\pT$ dependence of the $v_2$--$v_5$ in several $\nchrec$ event classes. The $v_5$ measurement is available only for three event-activity classes in a limited $\pT$ range. All flow harmonics show similar trends, i.e. they increase with $\pT$ up to 3--5 GeV and then decrease, but remain positive at higher $\pT$. For all event classes, the magnitude of the $v_n$ is largest for $n=2$, and decreases quickly with increasing $n$. The ATLAS data are compared to the measurement by the CMS experiment~\cite{Chatrchyan:2013nka} for an event-activity class in which the number of offline reconstructed tracks, $N_{\mathrm{trk}}^{\mathrm{off}}$, within $|\eta|<2.4$ and $\pT>0.4$ GeV is $220\leq N_{\mathrm{trk}}^{\mathrm{off}}<260$. This is comparable to the $220\leq\nchrec<260$ event class used in the ATLAS analysis. A similar recoil removal procedure, with $N_{\mathrm{trk}}^{\mathrm{off}}<20$ as the peripheral events, has been used for the CMS data. Excellent agreement is observed between the two results.

\begin{figure}[!h]
\centering
\includegraphics[width=0.9\columnwidth]{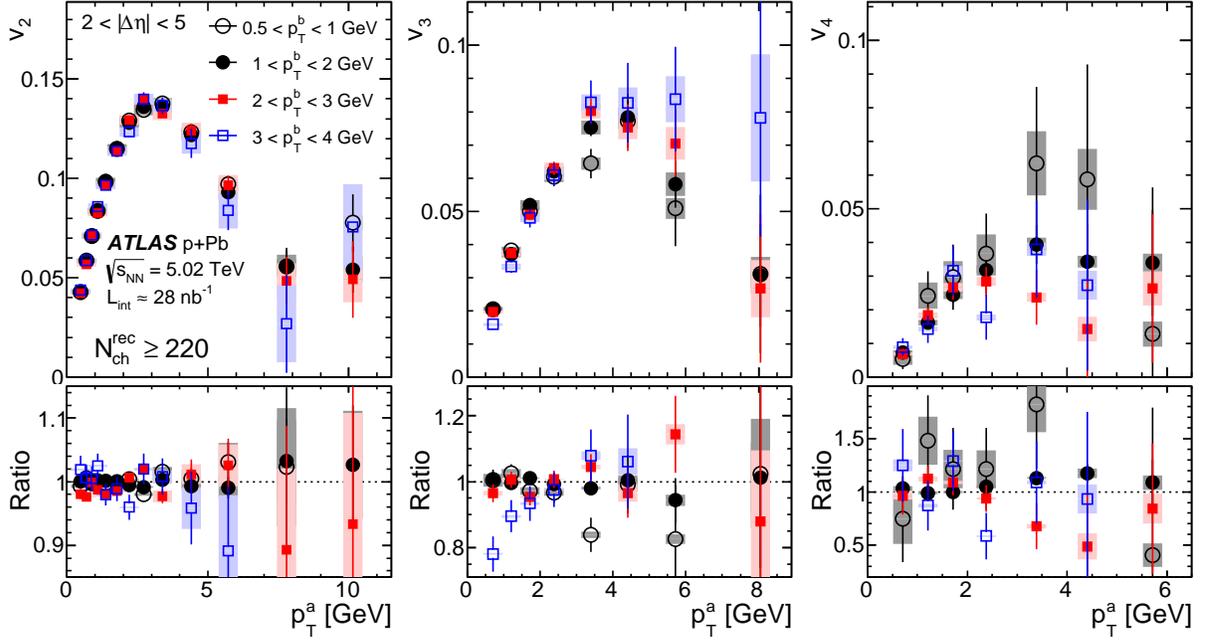}
\caption{\label{fig:resb3} The $v_2$ (left column), $v_3$ (middle column), and $v_4$ (right column) as a function of $\pta$ extracted using four $\ptref$ bins in the long-range region $|\Deta|>2$ for events with $\nchrec\geq220$. The ratio of the $v_n(\pta)$ in each $\ptref$ bin to those obtained with the default reference $\ptref$ bin of 1--3 GeV are shown in the bottom part of each column. The error bars and shaded bands represent the statistical and systematic uncertainties, respectively. }
\end{figure}

\begin{figure}[!h]
\centering
\includegraphics[width=0.9\columnwidth]{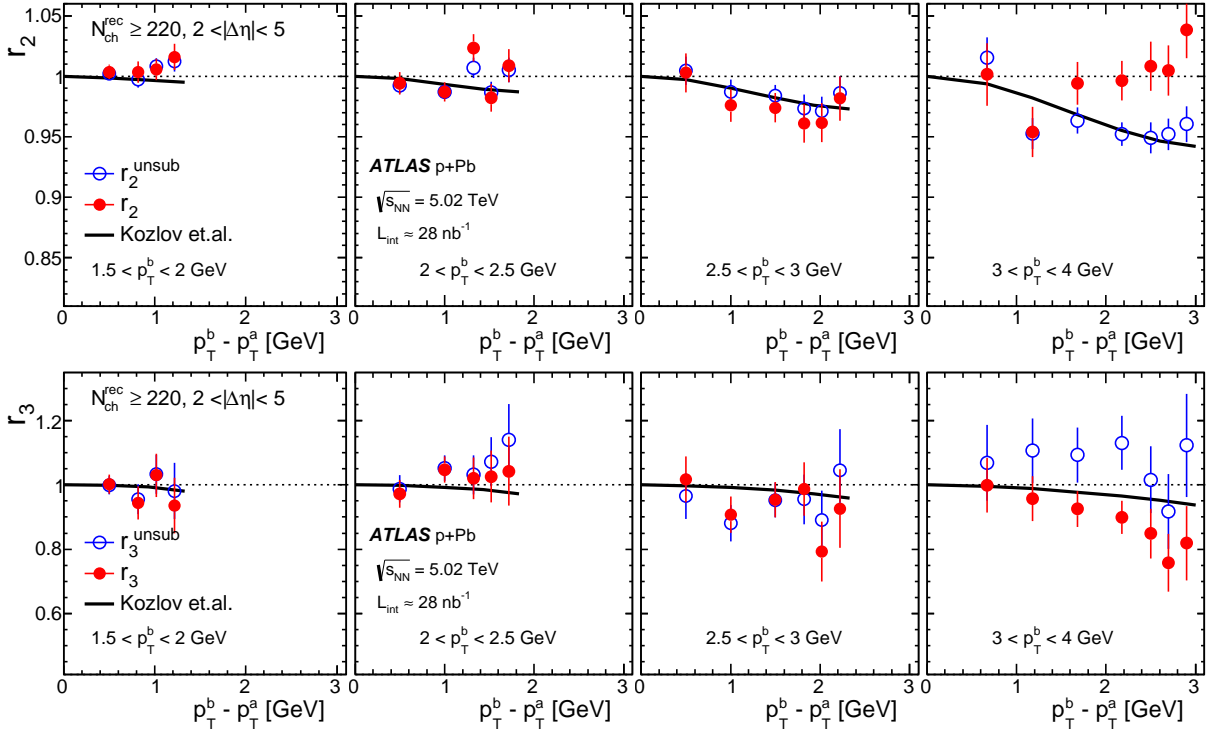}
\vspace*{-0.2cm}
\caption{\label{fig:rn} The values of factorization variable defined by Eq.~(\ref{eq:rn}) before (denoted by $r_n^{\rm{unsub}}$) and after (denoted by $r_n$) the subtraction of the recoil component. They are shown for $n=2$ (top row) and $n=3$ (bottom row) as a function of $\ptb-\pta$ in various $\ptb$ ranges for events in $\nchrec\geq220$. The solid lines represent a theoretical prediction from Ref.~\cite{Kozlov:2014fqa}. The error bars represent the total experimental uncertainties.}
\end{figure}

The extraction of the $v_n$ from $v_{n,n}$ relies on the factorization relation in Eq.~(\ref{eq:four2}). This factorization is checked by calculating $v_n$ using different ranges of $\ptb$ for events with $\nchrec\geq220$ as shown in Fig.~\ref{fig:resb3}. The factorization behavior can also be studied via the ratio~\cite{Gardim:2012im,Heinz:2013bua}
\begin{eqnarray} 
\label{eq:rn}
r_n(\pta,\ptb) = \frac{v_{n,n}(\pta,\ptb)}{\sqrt{v_{n,n}(\pta,\pta)v_{n,n}(\ptb,\ptb)}}\;,
\end{eqnarray}
with $r_n=1$ for perfect factorization. The results with recoil subtraction ($r_n$) and without subtraction ($r_n^{\mathrm{unsub}}$) are summarized in Fig.~\ref{fig:rn}, and they are shown as a function of $\ptb-\pta$, because by construction the ratios equal one for $\ptb=\pta$. This second method is limited to $\ptab\lesssim4$ GeV, since requiring both particles to be at high $\pT$ reduces the number of the available pairs for $v_{n,n}(\pta,\pta)$ or $v_{n,n}(\ptb,\ptb)$. In contrast, for the results shown in Fig.~\ref{fig:resb3}, using Eqs.~(\ref{eq:four2}) and (\ref{eq:four3}), the restriction applies to only one of the particles, i.e. $\ptb\lesssim4$ GeV.

Results in Figs.~\ref{fig:resb3} and \ref{fig:rn} show that, in the region where the statistical uncertainty is small, the factorization holds to within a few percent for $v_2$ over $0.5<\ptab<4$ GeV, within 10\% for $v_3$ over $0.5<\ptab<3$ GeV, and within 20--30\% for $v_4$ over $0.5<\ptab<4$ GeV (Fig.~\ref{fig:resb3} only). Furthermore, in this $\pT$ region, the differences between $r_n$ and $r_n^{\mathrm{unsub}}$ are very small ($<10\%$) as shown by Fig.~\ref{fig:rn}, consistent with the observation in Fig.~\ref{fig:resb1}. This level of factorization is similar to what was observed in peripheral Pb+Pb collisions~\cite{Aad:2012bu}.

Figure~\ref{fig:rn} also compares the $r_n$ data with a theoretical calculation from a viscous hydrodynamic model~\cite{Kozlov:2014fqa}. The model predicts at most a few percent deviation of $r_n$ from one, which is attributed to $\pT$-dependent decorrelation effects associated with event-by-event flow fluctuations~\cite{Gardim:2012im}. In most cases, the data are consistent with the prediction within uncertainties.
 
\begin{figure}[!h]
\centering
\includegraphics[width=1\columnwidth]{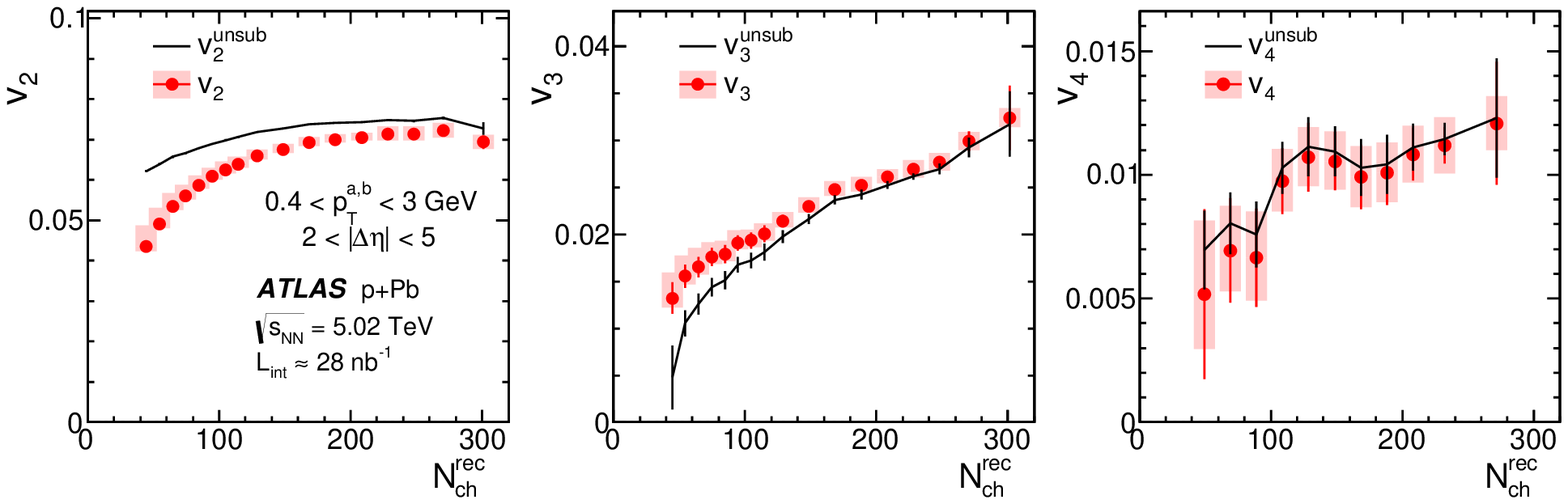}\\
\includegraphics[width=1\columnwidth]{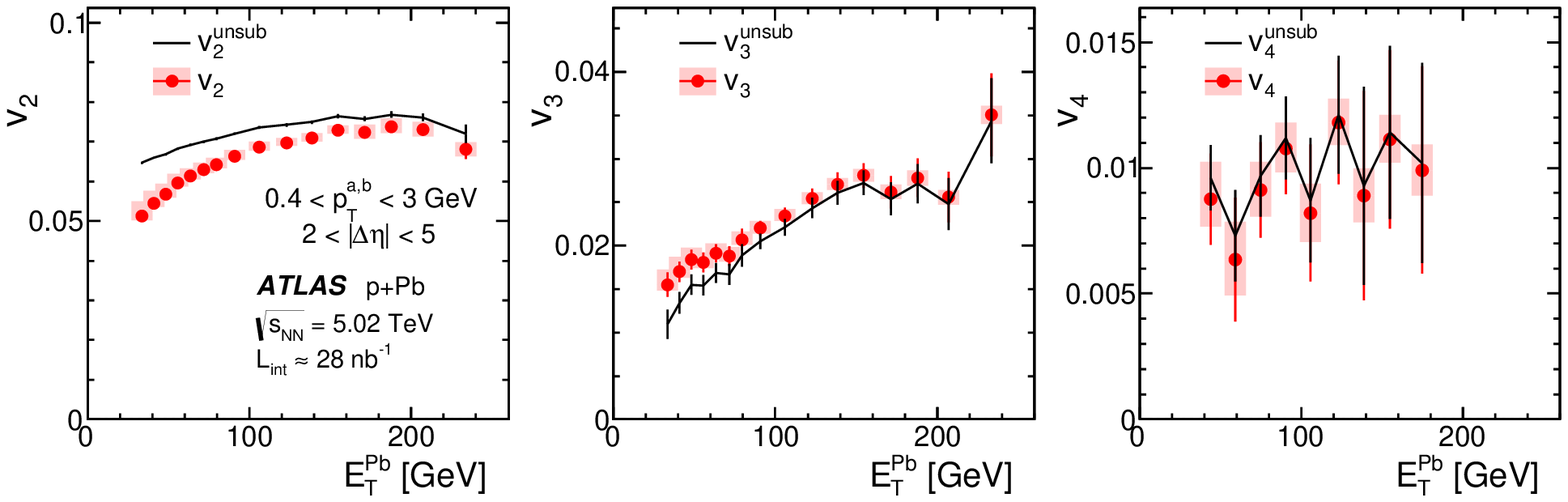}
\caption{\label{fig:resb2} The centrality dependence of $v_2$, $v_3$, and $v_4$ as a function of $\nchrec$ (top row) and $\ETfcal$ (bottom row) for pairs with $0.4<\ptab<3$ GeV and $|\Deta|>2$. The results are obtained with (symbols) and without (lines) the subtraction of the recoil contribution. The error bars and shaded boxes on $v_n$ data represent the statistical and systematic uncertainties, respectively, while the error bars on the $\vndir$ represent the combined statistical and systematic uncertainties.}
\end{figure}

Figure~\ref{fig:resb2} shows the centrality dependence of $v_2$, $v_3$, and $v_4$ as a function of $\nchrec$ and $\ETfcal$. The results are obtained for $0.4<\ptab<3$ GeV, both before and after subtraction of the recoil contribution. The difference between $\vndir$ and $v_n$ is very small in central collisions, up to 3--4\% for both event-activity definitions. For more peripheral collisions, the difference is larger and reaches 20--30\% for $\nchrec\sim40$ or $\ETfcal\sim30$ GeV. The sign of the difference also alternates in $n$ (already seen in Fig.~\ref{fig:resb1}): i.e. $\vndir>v_n$ for even $n$ and $\vndir<v_n$ for odd $n$. This behavior is characteristic of the influence of the away-side dijet contribution to $\vndir$.

The $v_n$ values in Fig.~\ref{fig:resb2} exhibit modest centrality dependence. The change of $v_2$ is less than 8\% over $140<\nchrec<300$ (top 0.5\% of MB-triggered events) or $130<\ETfcal<240$ GeV (top 0.05\% of MB-triggered events), covering about half of the full dynamic range. The centrality dependence of $v_3$ is stronger and exhibits a nearly linear increase with $\nchrec$ and $\ETfcal$.
 
\begin{figure}[!h]
\centering
\includegraphics[width=0.8\columnwidth]{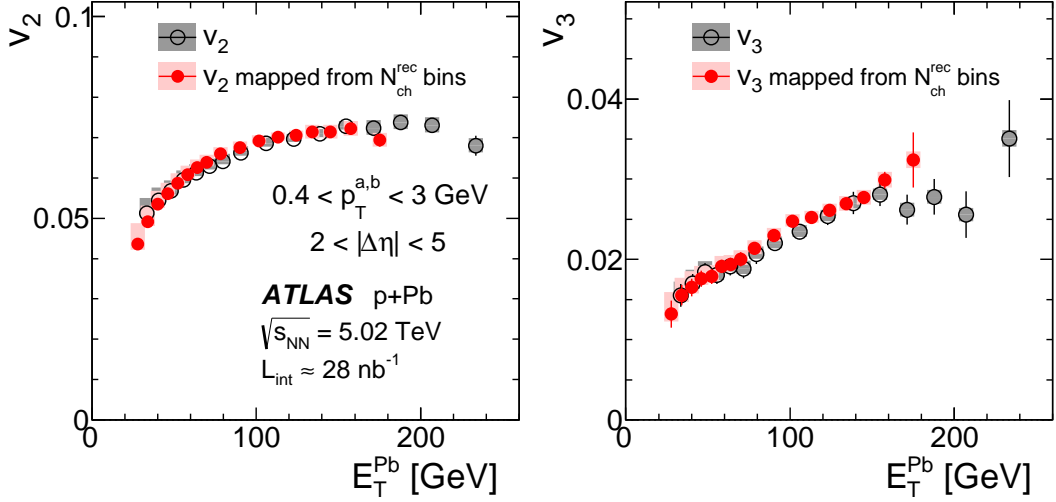}
\caption{\label{fig:resc1} The $v_2$ (left panel) and $v_3$ (right panel) as a function of $\ETfcal$ calculated directly for narrow ranges in $\ETfcal$ (open cicles) or obtained indirectly by mapping from the $\nchrec$-dependence of $v_n$ using the correlation data shown in Fig.~\ref{fig:2b}(b) (solid circles). The error bars and shaded boxes represent the statistical and systematic uncertainties, respectively.}
\end{figure}

Figure~\ref{fig:resb2} shows that the overall centrality dependence is similar for $\nchrec$ and $\ETfcal$. The correlation data (not the fit, Eq.~(\ref{eq:map})) in Fig.~\ref{fig:2b} are used to map the $\nchrec$-dependence in the top row of Fig.~\ref{fig:resb2} to a corresponding $\ETfcal$-dependence. The $\ETfcal$-dependence of $v_n$ mapped from $\nchrec$-dependence is then compared to the directly measured $\ETfcal$-dependence in Fig.~\ref{fig:resc1}. Good agreement is seen for $v_2$ and $v_3$.

\subsection{First-order Fourier coefficient $v_1$}
\label{sec:v1}
\begin{figure}[!h]
\centering
\includegraphics[width=1\columnwidth]{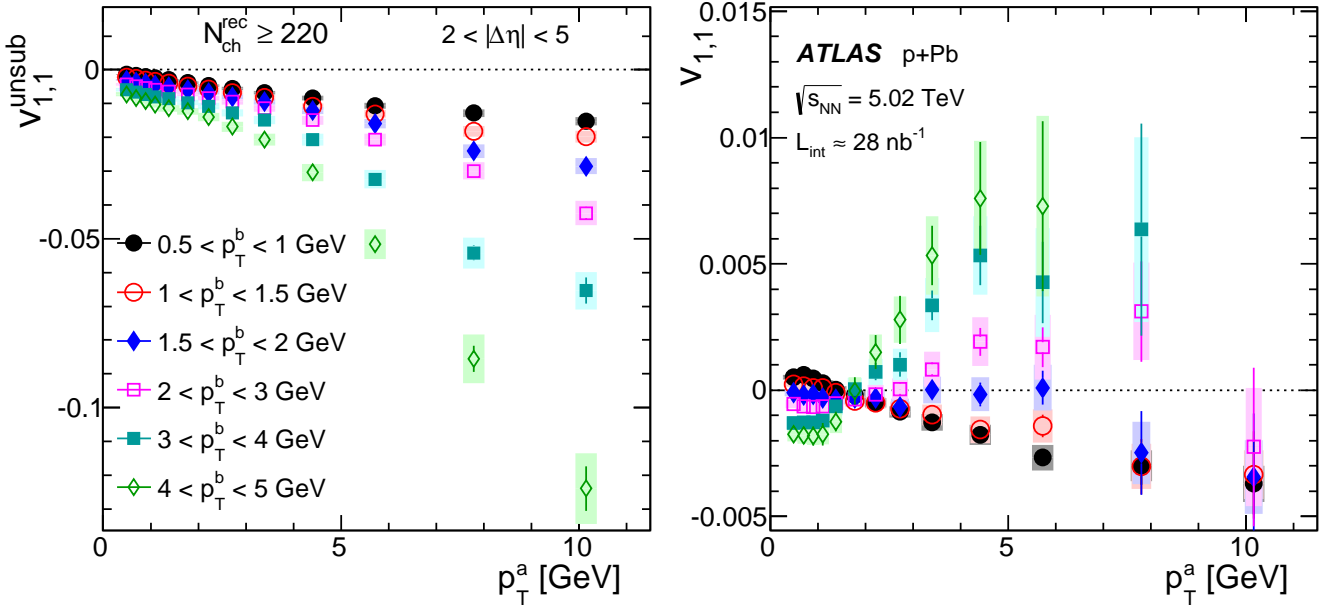}
\caption{\label{fig:v1a} The first-order harmonic of 2PC before recoil subtraction $v_{1,1}^{\mathrm{unsub}}$ (left panel) and after recoil subtraction $v_{1,1}$ (right panel) as a function of $\pta$ for different $\ptb$ ranges for events with $\nchrec\geq220$. The error bars and shaded boxes represent the statistical and systematic uncertainties, respectively.}
\end{figure}

A similar analysis is performed to extract the dipolar flow $v_1$. Figure~\ref{fig:v1a} shows the $v_{1,1}$ values as a function of $\pta$ in various ranges of $\ptb$ before and after the recoil subtraction. Before the recoil subtraction, $v_{1,1}^{\mathrm{unsub}}$ values are always negative and decrease nearly linearly with $\pta$ and $\ptb$, except for the $\pT$ region around 3--4 GeV where a shoulder-like structure is seen. This shoulder is very similar to that observed in A+A collisions, which is understood as a combined contribution from the negative recoil and positive dipolar flow in this $\pT$ range~\cite{Retinskaya:2012ky,Aad:2012bu} according to the following form~\cite{Borghini:2000cm,Borghini:2002mv}:
\begin{eqnarray}
\label{eq:v1}
v_{1,1}^{\mathrm{unsub}}(\pT^{\mathrm a},\pT^{\mathrm b}) \approx v_1(\pT^{\mathrm a})v_1(\pT^{\mathrm b})-\frac{\pT^{\mathrm a}\pT^{\mathrm b}}{M\langle \pT^2\rangle}\;,
\end{eqnarray}
where $M$ and $\langle \pT^2\rangle$ are the multiplicity and average squared transverse momentum of the particles in the whole event, respectively. The negative correction term reflects the global momentum conservation contribution, which is important in low-multiplicity events and at high $\pT$. The shoulder-like structure in Fig.~\ref{fig:v1a} reflects the contribution of the dipolar flow term $v_1(\pT^{\mathrm a})v_1(\pT^{\mathrm b})$. 

After the recoil subtraction, the magnitude of $v_{1,1}$ is greatly reduced, suggesting that most of the momentum conservation contribution has been removed. The resulting $v_{1,1}$ values cross each other at around $\pta\sim1.5$--2.0~GeV. This behavior is consistent with the expectation that the $v_1(\pT)$ function crosses zero at $\pT\sim1$--2~GeV, a feature that is also observed in A+A collisions~\cite{Retinskaya:2012ky,Aad:2012bu}. The trigger $\pT$ dependence of $v_1$ is obtained via a factorization procedure very similar to that discussed in Sec.~\ref{sec:fourier}
\begin{eqnarray}
\label{eq:v1b}
v_1(\pta) \equiv \frac{v_{1,1}(\pta,\ptref)}{v_1(\ptref)}\;,
\end{eqnarray}
where the dipolar flow in the associated $\pT$ bin, $v_1(\ptref)$, is defined as 
\begin{eqnarray}
\label{eq:v1c}
 v_{1}(\ptref)  = \mathrm{sign}(\ptref-\pT^0)\sqrt{\left|v_{1,1}(\ptref,\ptref)\right|}\;, 
\end{eqnarray}
where $\mathrm{sign}(\ptb-\pT^0)$ is the sign of the $v_{1}$, defined to be negative for $\ptb<\pT^0=1.5$ GeV and positive otherwise. This function is necessary in order to account for the sign change of $v_1$ at low $\pT$. 

To obtain the $v_1(\pta)$, three reference $\ptref$ ranges, 0.5--1 GeV, 3--4 GeV and 4--5 GeV, are used to first calculate $v_{1}(\ptref)$. These values are then inserted into Eq.~(\ref{eq:v1b}) to obtain three $v_1(\pta)$ functions. The uncertainties on the $v_1(\pta)$ values are calculated via an error propagation through Eqs.~(\ref{eq:v1b}) and (\ref{eq:v1c}). The calculation is not possible for $\ptb$ in the range of 1--3 GeV, where the $v_{1,1}$ values are close to zero and hence the resulting $v_{1}(\ptref)$ have large uncertainties.

\begin{figure}[!h]
\centering
\includegraphics[width=0.7\columnwidth]{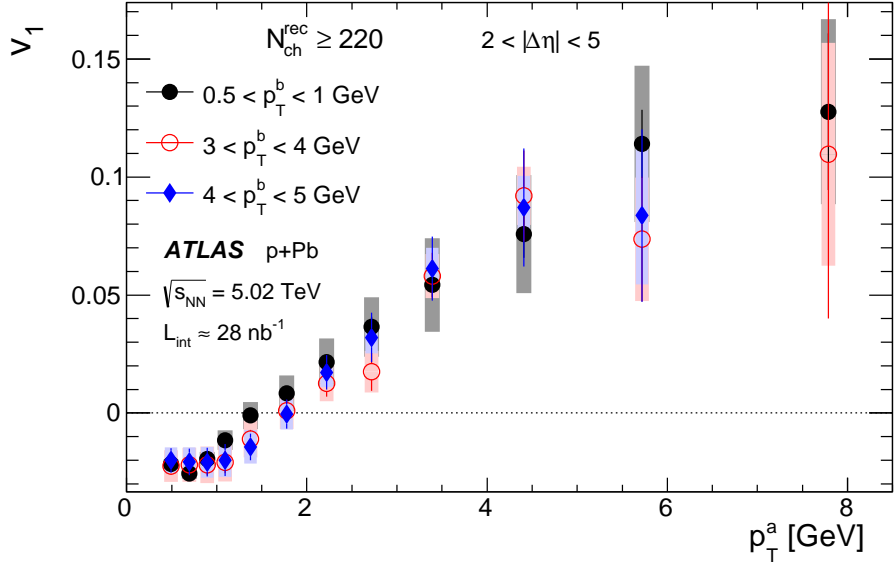}
\caption{\label{fig:v1b} The $\pta$ dependence of $v_{1}$ extracted using the factorization relations Eqs.~(\ref{eq:v1b}) and (\ref{eq:v1c}) in three reference $\ptref$ ranges for events with $\nchrec\geq220$. The error bars and shaded boxes represent the statistical and systematic uncertainties, respectively.}
\end{figure}

The results for $v_1(\pta)$ are shown in Fig.~\ref{fig:v1b} for these three reference $\ptref$ bins. They are consistent with each other. The $v_1$ value is negative at low $\pT$, crosses zero at around $\pT\sim1.5$ GeV, and increases to 0.1 at 4--6 GeV. This $\pT$-dependence is similar to the $v_1(\pT)$ measured by ATLAS experiment in Pb+Pb collisions at $\sqrtsNN=2.76$ TeV~\cite{Aad:2012bu}, except that the $v_1$ value in Pb+Pb collisions crosses zero at lower $\pT$ ($\sim 1.1$ GeV), which reflects the fact that the $\langle\pT\rangle$ in Pb+Pb at $\sqrtsNN=2.76$ TeV is smaller than that in $\pPb$ at $\sqrtsNN=5.02$ TeV.

\subsection{Comparison of $v_n$ results between high-multiplicity p+Pb and peripheral Pb+Pb collisions}
In the highest multiplicity $\pPb$ collisions, the charged particle multiplicity, $\nchrec$, can reach more than 350 in $|\eta|<2.5$ and $\ETfcal$ close to 300 GeV on the Pb-fragmentation side. This activity is comparable to Pb+Pb collisions at $\sqrtsnn =2.76$~TeV in the 45--50\% centrality interval, where the long-range correlation is known to be dominated by collective flow. Hence a comparison of the $v_n$ coefficients in similar event activity for the two collision systems can improve our current understanding of the origin of the long-range correlations.

\begin{figure}[!h]
\centering
\includegraphics[width=1\columnwidth]{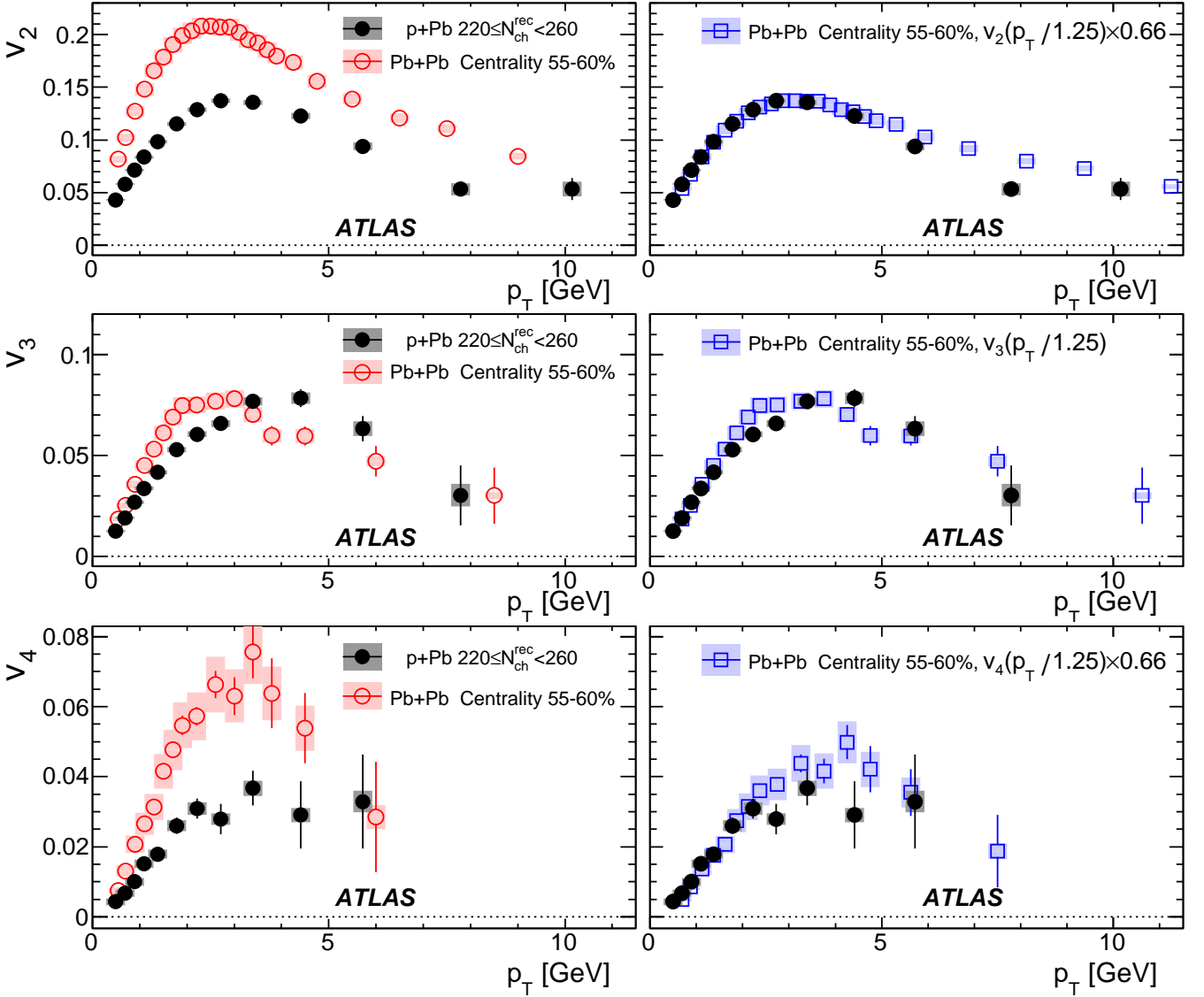}
\caption{\label{fig:comp1} The coefficients $v_2$ (top row), $v_3$ (middle row) and $v_4$ (bottom row) as a function of $\pT$ compared between $\pPb$ collisions with $220\leq\nchrec<260$ in this analysis and Pb+Pb collisions in 55--60\% centrality from Ref.~\cite{Aad:2012bu}. The left column shows the original data with their statistical (error bars) and systematic uncertainties (shaded boxes). In the right column, the same Pb+Pb data are rescaled horizontally by a constant factor of 1.25, and the $v_2$ and $v_4$ are also down-scaled by an empirical factor of 0.66 to match the $\pPb$ data.}
\end{figure}

The left column of Fig.~\ref{fig:comp1} compares the $v_n$ values from $\pPb$ collisions with $220\leq\nchrec<260$ to the $v_n$ values for Pb+Pb collisions in the 55--60\% centrality interval from Ref.~\cite{Aad:2012bu}. These two event classes are chosen to have similar efficiency-corrected multiplicity of charged particles with $\pT>0.5$~GeV and $|\eta|<2.5$, characterized by its average value (\nchave) and its standard deviation ($\sigma$): $\nchave\pm\sigma\approx259\pm13$ for $\pPb$ collisions and $\nchave\pm\sigma\approx 241\pm43$ for Pb+Pb collisions.

The Pb+Pb results on $v_n$~\cite{Aad:2012bu} were obtained via an event-plane method by correlating tracks in $\eta>0$ ($\eta < 0$) with the event plane determined in the FCal in the opposite hemisphere. The larger $v_2$ values in Pb+Pb collisions can be attributed to the elliptic collision geometry of the Pb+Pb system, while the larger $v_4$ values are due to the non-linear coupling between $v_2$ and $v_4$ in the collective expansion~\cite{Luzum:2010ae}. The $v_3$ data for Pb+Pb collisions are similar in magnitude to those in $\pPb$ collisions. However, the $\pT$ dependence of $v_n$ is different for the two systems. These observations are consistent with similar comparisons performed by the CMS experiment~\cite{Chatrchyan:2013nka}.

Recently, Basar and Teaney~\cite{Basar:2013hea} have proposed a method to rescale the Pb+Pb data for a 
proper comparison to the $\pPb$ data. They argue that the $v_n(\pT)$ shape in the two collision systems are related to each other by a constant scale factor of $K=1.25$ accounting for the difference in their $\langle\pT\rangle$, and that one should observe a similar $v_n(\pT)$ dependence shape after rescaling the $\pT$ measured in Pb+Pb collisions to get $v_n(\pT/K)$. The difference in the overall magnitude of $v_2$ after the $\pT$-rescaling is entirely due to the elliptic geometry of Pb+Pb collisions.

In order to test this idea, the $\pT$ for Pb+Pb collisions are rescaled by the constant factor of 1.25 and $v_n$ values with rescaled $\pT$ are displayed in the right column of Fig.~\ref{fig:comp1}. Furthermore, the magnitudes of $v_2$ and $v_4$ are also rescaled by a common empirical value of 0.66 to approximately match the magnitude of the corresponding $\pPb$ $v_n$ data. The rescaled $v_n$ results are shown in the right column and compared to the $\pPb$ $v_n$ data. They agree well with each other, in particular in the low-$\pT$ region ($\pT<2$--4 GeV) where the statistical uncertainties are small.

\section{Summary}
\label{sec:summary}
This paper presents measurements of two-particle correlation (2PC) functions and the first five azimuthal harmonics $v_1$--$v_5$ in $\sqrtsnn =5.02$~TeV $\pPb$ collisions with a total integrated luminosity of approximately 28~$\mathrm{nb}^{-1}$ recorded by the ATLAS detector at the LHC. The two-particle correlations and $v_n$ coefficients are obtained as a function of $\pT$ for pairs with $2<|\Delta \eta|<5$ in different intervals of event activity, defined by either $\nchrec$, the number of reconstructed tracks with $\pT>0.4$ GeV and $|\eta|<2.5$, or $\ETfcal$, the total transverse energy over $-4.9<\eta<-3.2$ on the Pb-fragmentation side.

Significant long-range correlations (extending to $|\Deta|=5$) are observed for pairs at the near-side ($|\Dphi|<\pi/3$) over a wide range of transverse momentum ($\pT<12$ GeV) and broad ranges of $\nchrec$ and $\ETfcal$. A similar long-range correlation is also observed on the away-side ($|\Dphi|>2\pi/3$), after subtracting the recoil contribution estimated using the 2PC in low-activity events. 

The azimuthal structure of these long-range correlations is quantified using the Fourier coefficients $v_2$--$v_5$ as a function of $\pT$. The $v_n$ values increase with $\pT$ to 3--4 GeV and then decrease for higher $\pT$, but remain positive in the measured $\pT$ range. The overall magnitude of $v_n(\pT)$ is observed to decrease with $n$. The magnitudes of $v_n$ also increase with both $\nchrec$ and $\ETfcal$. The $v_2$ values seem to saturate at large $\nchrec$ or $\ETfcal$ values, while the $v_3$ values show a linear increase over the measured $\nchrec$ or $\ETfcal$ range. The first-order harmonic $v_1$ is also extracted from the 2PC. The $v_1(\pT)$ function is observed to change sign at $\pT\approx 1.5$--2.0 GeV and to increase to about 0.1 at $\pT>4$ GeV.

The extracted $v_2(\pT)$, $v_3(\pT)$ and $v_4(\pT)$ are compared to the $v_n$ coefficients in Pb+Pb collisions at $\sqrtsnn =2.76$~TeV with similar $\nchrec$. After applying a scale factor of $K=1.25$ that accounts for the difference of mean $\pT$ in the two collision systems as suggested in Ref.~\cite{Basar:2013hea}, the shape of the $v_n(\pT/K)$ distribution in Pb+Pb collision is found to be similar to the shape of $v_n(\pT)$ distribution in $\pPb$ collisions. This suggests that the long-range ridge correlations in high-multiplicity $\pPb$ collisions and peripheral Pb+Pb collisions are driven by similar dynamics.

\section*{ACKNOWLEDGEMENTS}
We thank CERN for the very successful operation of the LHC, as well as the
support staff from our institutions without whom ATLAS could not be
operated efficiently.

We acknowledge the support of ANPCyT, Argentina; YerPhI, Armenia; ARC,
Australia; BMWF and FWF, Austria; ANAS, Azerbaijan; SSTC, Belarus; CNPq and FAPESP,
Brazil; NSERC, NRC and CFI, Canada; CERN; CONICYT, Chile; CAS, MOST and NSFC,
China; COLCIENCIAS, Colombia; MSMT CR, MPO CR and VSC CR, Czech Republic;
DNRF, DNSRC and Lundbeck Foundation, Denmark; EPLANET, ERC and NSRF, European Union;
IN2P3-CNRS, CEA-DSM/IRFU, France; GNSF, Georgia; BMBF, DFG, HGF, MPG and AvH
Foundation, Germany; GSRT and NSRF, Greece; ISF, MINERVA, GIF, I-CORE and Benoziyo Center,
Israel; INFN, Italy; MEXT and JSPS, Japan; CNRST, Morocco; FOM and NWO,
Netherlands; BRF and RCN, Norway; MNiSW and NCN, Poland; GRICES and FCT, Portugal; MNE/IFA, Romania; MES of Russia and ROSATOM, Russian Federation; JINR; MSTD,
Serbia; MSSR, Slovakia; ARRS and MIZ\v{S}, Slovenia; DST/NRF, South Africa;
MINECO, Spain; SRC and Wallenberg Foundation, Sweden; SER, SNSF and Cantons of
Bern and Geneva, Switzerland; NSC, Taiwan; TAEK, Turkey; STFC, the Royal
Society and Leverhulme Trust, United Kingdom; DOE and NSF, United States of
America.

The crucial computing support from all WLCG partners is acknowledged
gratefully, in particular from CERN and the ATLAS Tier-1 facilities at
TRIUMF (Canada), NDGF (Denmark, Norway, Sweden), CC-IN2P3 (France),
KIT/GridKA (Germany), INFN-CNAF (Italy), NL-T1 (Netherlands), PIC (Spain),
ASGC (Taiwan), RAL (UK) and BNL (USA) and in the Tier-2 facilities
worldwide.

\bibliographystyle{atlasBibStyleWoTitle}
\providecommand{\href}[2]{#2}\begingroup\raggedright\endgroup
\onecolumngrid
\clearpage 
\begin{flushleft}
{\Large The ATLAS Collaboration}

\bigskip

G.~Aad$^{\rm 84}$,
B.~Abbott$^{\rm 112}$,
J.~Abdallah$^{\rm 152}$,
S.~Abdel~Khalek$^{\rm 116}$,
O.~Abdinov$^{\rm 11}$,
R.~Aben$^{\rm 106}$,
B.~Abi$^{\rm 113}$,
M.~Abolins$^{\rm 89}$,
O.S.~AbouZeid$^{\rm 159}$,
H.~Abramowicz$^{\rm 154}$
% ...
\bigskip
\\
% ...
$^{11}$ Institute of Physics, Azerbaijan Academy of Sciences, Baku, Azerbaijan\\
% ...
$^{84}$ CPPM, Aix-Marseille Universit{\'e} and CNRS/IN2P3, Marseille, France\\
% ...
$^{89}$ Department of Physics and Astronomy, Michigan State University, East Lansing MI, United States of America\\
% ...
$^{106}$ Nikhef National Institute for Subatomic Physics and University of Amsterdam, Amsterdam, Netherlands\\
% ...
$^{112}$ Homer L. Dodge Department of Physics and Astronomy, University of Oklahoma, Norman OK, United States of America\\
$^{113}$ Department of Physics, Oklahoma State University, Stillwater OK, United States of America\\
% ...
$^{116}$ LAL, Universit{\'e} Paris-Sud and CNRS/IN2P3, Orsay, France\\
% ...
$^{152}$ Institute of Physics, Academia Sinica, Taipei, Taiwan\\
% ...
$^{154}$ Raymond and Beverly Sackler School of Physics and Astronomy, Tel Aviv University, Tel Aviv, Israel\\
% ...
$^{159}$ Department of Physics, University of Toronto, Toronto ON, Canada\\
% ...
\end{flushleft}


\end{document}